\documentclass[preprints,entropy,article,accept,oneauthor,pdftex]{./mdpi} 

\usepackage{graphicx}
\usepackage{epstopdf}
\usepackage{rotating}
\usepackage{dsfont}
\usepackage{color}
\usepackage{tikz}
\usetikzlibrary{plotmarks}
\usepackage{pgfplots}
\usetikzlibrary{calc}
\usetikzlibrary{shapes,arrows}
\usetikzlibrary{decorations.markings}
\usetikzlibrary{positioning}
\pgfplotsset{compat=1.10}
\usetikzlibrary{calc}
\usetikzlibrary{shapes,arrows}
\usetikzlibrary{decorations.markings}

\usepackage[utf8]{inputenc}
\usepackage{authblk}

\usepackage{accents}
\makeatletter
\newcommand{\thickbar}{\mathpalette\@thickbar}
\newcommand{\@thickbar}[2]{{#1\mkern1.5mu\vbox{
			\sbox\z@{$#1\mkern-1.5mu#2\mkern-1.5mu$}%
			\sbox\tw@{$#1\overline{#2}$}%
			\dimen@=\dimexpr\ht\tw@-\ht\z@-.8\p@\relax
			\hrule\@height.8\p@ 
			\vskip\dimen@
			\box\z@}\mkern1.5mu}
}
\makeatother

\DeclareFontFamily{U}{mathx}{\hyphenchar\font45}
\DeclareFontShape{U}{mathx}{m}{n}{<-> mathx10}{}
\DeclareSymbolFont{mathx}{U}{mathx}{m}{n}
\DeclareMathAccent{\widebar}{0}{mathx}{"73}

\newcommand{\Enc}{\mathsf{Enc}}
\newcommand{\Dec}{\mathsf{Dec}}

\firstpage{1} 
\pubvolume{1}
\issuenum{1}
\articlenumber{0}
\pubyear{2021}
\copyrightyear{2020}
\datereceived{} 
\dateaccepted{} 
\datepublished{} 
\hreflink{https://doi.org/} 

\Title{Function Computation Under Privacy, Secrecy, Distortion, and Communication Constraints}

\TitleCitation{Function Computation Under Privacy, Secrecy, Distortion, and Communication Constraints}

\Author{Onur G{\"u}nl{\"u}\orcidA{}}

\AuthorNames{Onur G{\"u}nl{\"u}}

\AuthorCitation{G{\"u}nl{\"u}, O.}

\address{%
Chair of Communications Engineering and Security, University of Siegen, 57076 Siegen, Germany; onur.guenlue@uni-siegen.de
}

\corres{Correspondence:onur.guenlue@uni-siegen.de}

\conference{the 2021 International ITG Workshop on Smart Antennas in \cite{ourWSA2021LossyMultiObservationFunctionComp} and the 2021 Asilomar Conference on Signals, Systems, and Computers in \cite{ourAsilomarSecurePrivateFunct}} 

\abstract{The problem of reliable function computation is extended by imposing privacy, secrecy, and storage constraints on a remote source whose noisy measurements are observed by multiple parties. The main additions to the classic function computation problem include 1) privacy leakage to an eavesdropper is measured with respect to the remote source rather than the transmitting terminals' observed sequences; 2) the information leakage to a fusion center with respect to the remote source is considered as a new privacy leakage metric; 3) the function computed is allowed to be a distorted version of the target function, which allows to reduce the storage rate as compared to a reliable function computation scenario in addition to reducing secrecy and privacy leakages; 4) two transmitting node observations are used to compute a function. Inner and outer bounds on the rate regions are derived for lossless and lossy single-function computation with two transmitting nodes, which recover previous results in the literature. For special cases, including invertible and partially invertible functions, and degraded measurement channels, simplified lossless and lossy rate region bounds are established, and one region is evaluated as an example scenario.}


\begin{document}
\section{Introduction} \label{sec:intro}
We consider function computation scenarios in a network with multiple nodes involved. Each node observes a random sequence and all observed random sequences are modeled to be correlated. Recent advancements in network function virtualization \cite{NFVReview} and distributed machine learning applications \cite{DistributedlearningReview} make function computation in a wireless network via software defined networking an important practical problem that should be tackled to improve the performance of future communication systems. In a classic function computation scenario, the nodes exchange messages through authenticated, noiseless, and public communication links, which results in undesired information leakage about the function computed \cite{YaoSecureFunctionComp,YaoSecureFunctionComp2,TyagiNarayanSecureCompute}. Furthermore, it is possible to reduce the amount of public communications \cite{CodingforComputing,OurJSAITTutorial} by using distributed lossless or lossy source coding methods; see \cite{IshwarCompute,TchamkertenCompute,GastparCompute,KumarCompute,ViswanathCompute} for several extensions. The former method uses Slepian-Wolf (SW) coding \cite{SW} constructions and the latter allows the function computed to be a distorted version of the target function and applies Wyner-Ziv (WZ) coding \cite{WZCard} methods that result in further reductions compared to the former. A decrease in public communication is important also to limit the information about the computed function leaked to an eavesdropper in the same network, i.e., \emph{secrecy leakage}. In addition to the public messages, an eavesdropper has generally access to a random sequence correlated with other sequences; see \cite{GoldenbaumComp,TyagiWatanabeTIT,PrabhakaranCompute} for various secure function computation extensions.

An important addition to the secure function computation model is a \emph{privacy} constraint that measures the amount of information about the observed sequence leaked to an eavesdropper \cite{LifengFCTrans}. Providing privacy is necessary to ensure confidentiality of a private sequence that can be reused for future function computations \cite{bizimMMMMTIFS,benimdissertation}. An extension of the results in \cite{LifengFCTrans} are given in \cite{ourISITSecurePrivateFunctArxiv}, where two privacy constraints are considered on a remote source whose different noisy measurements are observed by multiple nodes in the same network. The extension in \cite{ourISITSecurePrivateFunctArxiv} is different from the previous secure and private function computation models due to the posit that there exists a remote source that is the main reason for the correlation between the random sequences observed by the nodes in the same network. It is illustrated via practical examples that considering a remote source hinders unexpected decrease in reliability and unnoticed secrecy leakage \cite{benimdissertation}. Similarly, such a remote source model is proposed, e.g., in \cite{groundtruthauthentication} for biometric secrecy and in \cite{bizimKittipongTIFS, bizimITW} for user or device authentication problems. It is shown in \cite{ourISITSecurePrivateFunctArxiv} that with such a remote source model two different privacy leakage rate values should be limited, unlike a single constraint considered in \cite{LifengFCTrans}.

We consider a private remote source whose three noisy versions are used for secure single-function computation. Suppose two nodes transmit public indices to a fusion center to compute one function. In \cite{ourISITSecurePrivateFunctArxiv}, for each function computation one node sends a public  index to a fusion center. In \cite{LifengFCTrans}, cases with two transmitting nodes for function computation are considered for a visible source model, whose results are improved in this work for a remote source model with an additional privacy leakage constraint. Furthermore, we also consider function computation scenarios where the function computed is allowed to be a distorted version of the target function, which is relevant for various recent function computation applications.

\subsection{Models for Function Inputs and Outputs}
We consider noisy remote source output measurements that are independent and identically distributed (i.i.d.) according to a fixed probability distribution and that are inputs of a target function. This model is reasonable if, e.g., one uses transform-coding algorithms from \cite{MLbasedTransform,bizimMDPI,SlavMostRelevant,Campisi} to extract almost i.i.d. symbols, as applied in the biometric security, physical unclonable function, and image and video coding literature. Furthermore, the set of target functions we study are applied per-letter, i.e., the same function is applied to each input symbol; see Section~\ref{sec:systemmodel} below. These functions are realistic and are used in various recent applications, such as distributed and federated learning applications where the same loss function is applied to each data example \cite{FederatedLearningFirst}.

\subsection{Summary of Contributions}
We extend the lossless and lossy rate region analysis of the single-function computation model with one transmitting node in \cite{ourISITSecurePrivateFunctArxiv} to consider two transmitting nodes with joint secrecy and privacy constraints, as well as a distortion constraint on the computed function. A summary of the main contributions is as follows.
\begin{itemize}
	\item The lossless single-function computation model with two transmitting nodes is considered and an inner bound for the rate region that characterizes the optimal trade-off between secrecy, privacy, storage, and distortion constraints is established by using the output statistics of random binning (OSRB) method \cite{AminOSRB}. An outer bound for the same rate region is also provided by using standard properties of Shannon entropy. Inner and outer bounds are shown to not match in general due to different Markov chains imposed.
	\item The proposed inner and outer bounds are extended for the lossy single-function computation model with two transmitting nodes by considering a distortion metric. Furthermore, effects of considering a distortion constraint, rather than a reliability constraint, on the function computation are discussed.
	\item  For both partially invertible functions, which define a set that is a proper superset of the set of invertible functions, and invertible functions, we establish simplified lossless and lossy rate region bounds.
	\item The simplified rate region bounds for invertible functions are further simplified when the eavesdropper's measurement channel is physically degraded with respect to the fusion center's channel or vice versa, which results in different bounds on the rates.
	\item We evaluate an achievable rate region for a physically degraded case with multiplicative Bernoulli noise components.
\end{itemize}

\subsection{Organization}
This paper is organized as follows. In Section~\ref{sec:systemmodel}, we introduce the lossless and lossy single-function computation problems with two transmitting nodes under secrecy, privacy, storage, and reliability or distortion constraints. In Section~\ref{sec:innerouterboundsforgeneral}, we present the inner and outer bounds for the rate regions of the introduced problems and discuss that the bounds differ because of different Markov chains imposed. In Section~\ref{sec:specialcases}, we establish simplified lossless and lossy rate region bounds for invertible  functions, partially invertible functions, and two different degraded measurement channels, and an achievable rate region for an example case is evaluated. In Section~\ref{sec:proofinnerouter2lossless}, we  offer proofs of the inner and outer bounds for the lossless single-function computations with two transmitting nodes. In Section~\ref{sec:conclusion}, we conclude the paper.

\subsection{Notation}
Upper case letters represent random variables and lower case letters their realizations. A superscript denotes a sequence of variables, e.g., $\displaystyle X^n\!=\!X_1,X_2,\ldots, X_i,\ldots, X_n$, and a subscript $i$ denotes the position of a variable in a sequence. A random variable $\displaystyle X$ has probability distribution $\displaystyle P_X$. Calligraphic letters such as $\displaystyle \mathcal{X}$ denote sets, set sizes are written as $\displaystyle |\mathcal{X}|$. Given any $a\in\mathbb{R}$, define $[a]^-=\min\{a,0\}$. $H_b(c)\!=\!-c\log_2 c- (1\!-\!c)\log_2 (1\!-\!c)$ is the binary entropy function for any $c\in [0,1]$.

\section{System Model}\label{sec:systemmodel}
\begin{figure}
	\centering
	\resizebox{0.75\linewidth}{!}{
		\begin{tikzpicture}
		\node (so) at (-1.5,-3.3) [draw,rounded corners = 5pt, minimum width=0.8cm,minimum height=0.8cm, align=left] {$P_X$};
		\node (a) at (0,-0.5) [draw,rounded corners = 6pt, minimum width=2.2cm,minimum height=0.8cm, align=left] {$W_1 = \Enc_1(\widetilde{X}_1^n)$};
		\node (c) at (5,-3.3) [draw,rounded corners = 5pt, minimum width=1.3cm,minimum height=0.6cm, align=left] {$P_{YZ|X}$};
		\node (f) at (0,-2.05) [draw,rounded corners = 5pt, minimum width=1cm,minimum height=0.6cm, align=left] {$P_{\widetilde{X}_1|X}$};
		\node (b) at (5,-0.5) [draw,rounded corners = 6pt, minimum width=2.2cm,minimum height=0.8cm, align=left] {$\widehat{f^n}= \Dec\left(W_1,W_2,Y^n\right)$};
		\node (g) at (5,-5) [draw,rounded corners = 5pt, minimum width=1cm,minimum height=0.6cm, align=left] {EVE};
		\draw[decoration={markings,mark=at position 1 with {\arrow[scale=1.5]{latex}}},
		postaction={decorate}, thick, shorten >=1.4pt] (a.east) -- (b.west) node [midway, above] {$W_1$};
		\node (a1) [below of = a, node distance = 2.8cm] {$X^n$};
		\draw[decoration={markings,mark=at position 1 with {\arrow[scale=1.5]{latex}}},
		postaction={decorate}, thick, shorten >=1.4pt] ($(c.north)+(0.0,0)$) -- ($(b.south)+(0.0,0)$) node [midway, right] {$Y^n$};
		\draw[decoration={markings,mark=at position 1 with {\arrow[scale=1.5]{latex}}},
		postaction={decorate}, thick, shorten >=1.4pt] (so.east) -- (a1.west);
		\draw[decoration={markings,mark=at position 1 with {\arrow[scale=1.5]{latex}}},
		postaction={decorate}, thick, shorten >=1.4pt] (a1.north) -- (f.south);
		\draw[decoration={markings,mark=at position 1 with {\arrow[scale=1.5]{latex}}},
		postaction={decorate}, thick, shorten >=1.4pt] (f.north) -- (a.south) node [midway, left] {$\widetilde{X}_1^n$};
		\draw[decoration={markings,mark=at position 1 with {\arrow[scale=1.5]{latex}}},
		postaction={decorate}, thick, shorten >=1.4pt, dashed] (a1.east) -- ($(c.west)-(0,0.0)$) node [above  left] {$X^n$};
		\draw[decoration={markings,mark=at position 1 with {\arrow[scale=1.5]{latex}}},
		postaction={decorate}, thick, shorten >=1.4pt] (c.south) -- (g.north) node [midway, right] {$Z^n$};
		\node (b2) [right of = b, node distance = 3cm] {$\widehat{f^n}$};
		\draw[decoration={markings,mark=at position 1 with {\arrow[scale=1.5]{latex}}},
		postaction={decorate}, thick, shorten >=1.4pt] (b.east) -- (b2.west);
		\draw[decoration={markings,mark=at position 1 with {\arrow[scale=1.5]{latex}}},
		postaction={decorate}, thick, shorten >=1.4pt] ($(a.east)+(1,0)$) -- ($(a.east)+(1,-4.35)$) -- ($(a.east)+(1,-4.35)$) -- ($(g.west)+(0,0.15)$) node [above left=0.15cm and 0.5cm of g.west] {$W_1$};
		\node (a2) at (0,-6.5) [draw,rounded corners = 6pt, minimum width=2.2cm,minimum height=0.8cm, align=left] {$W_2 = \Enc_2(\widetilde{X}_2^n)$};
		\node (f2) at (0,-5) [draw,rounded corners = 5pt, minimum width=1cm,minimum height=0.6cm, align=left] {$P_{\widetilde{X}_2|X}$};
		\draw[decoration={markings,mark=at position 1 with {\arrow[scale=1.5]{latex}}},
		postaction={decorate}, thick, shorten >=1.4pt] (a2.east) -- ($ (b2.south) -(2,5.65)$) -- ($(b.south)+(1.0,0)$)  node [midway, right] {$W_2$};
		\draw[decoration={markings,mark=at position 1 with {\arrow[scale=1.5]{latex}}},
		postaction={decorate}, thick, shorten >=1.4pt] (a1.south) -- (f2.north);
		\draw[decoration={markings,mark=at position 1 with {\arrow[scale=1.5]{latex}}},
		postaction={decorate}, thick, shorten >=1.4pt] (f2.south) -- (a2.north) node [midway, left] {$\widetilde{X}_2^n$};
		\draw[decoration={markings,mark=at position 1 with {\arrow[scale=1.5]{latex}}},
		postaction={decorate}, thick, shorten >=1.4pt] ($(a2.east)+(1,0)$) -- ($(a2.east)+(1,1.35)$) -- ($(a2.east)+(1,1.35)$) -- ($(g.west)+(0,-0.15)$) node [below left=0.15cm and 0.5cm of g.west] {$W_2$};
		\end{tikzpicture}
	}
	\caption{Single-function computation problem with two transmitting nodes under secrecy, privacy, and storage (or communication) constraints.
	}\label{fig:TIFSTwoTransmitHiddenFunction}
	\vspace*{-0.4cm}
\end{figure}
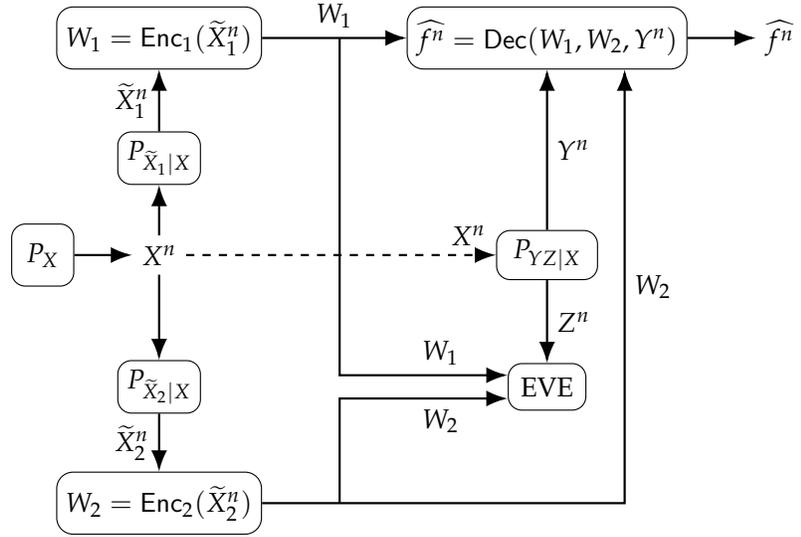

We consider the single-function computation model with two transmitting nodes illustrated in Figure~\ref{fig:TIFSTwoTransmitHiddenFunction}. Noisy measurements $\widetilde{X}_1^n$ and $\widetilde{X}_2^n$ of an i.i.d. remote source $X^n\sim P^n_X$ through memoryless channels $P_{\widetilde{X}_1|X}$ and $P_{\widetilde{X}_2|X}$, respectively, are observed by two legitimate nodes in a network. Similarly, other noisy measurements $Y^n$ and $Z^n$ of the same remote source are observed by the fusion center and eavesdropper (Eve), respectively, through another memoryless channel $P_{YZ|X}$. Encoders $\Enc_1(\cdot)$ and $\Enc_2(\cdot)$ of the legitimate nodes send indices $W_1$ and $W_2$, respectively, to the fusion center over public communication links with storage rate constraints. The fusion center decoder $\Dec(\cdot)$ then uses its observed noisy sequence $Y^n$ and the public indices $W_1$ and $W_2$ to estimate a function $f^n(\widetilde{X}_1^n,\widetilde{X}_2^n,Y^n)$ such that 
\begin{align}
f^n(\widetilde{X}_1^n,\widetilde{X}_2^n,Y^n) = {\{f(\widetilde{X}_{1,i},\widetilde{X}_{2,i},Y_i)\}}_{i=1}^n.
\end{align}
The source and measurement alphabets are finite sets.

A natural secrecy leakage constraint is to minimize the information leakage about the function output $f^n(\widetilde{X}_1^n,\widetilde{X}_2^n,Y^n)$  to eavesdropper. However, its analysis depends on the specific function $f(\cdot,\cdot,\cdot)$ computed, so we impose below another secrecy leakage constraint that does not depend on the function used and that provides an upper bound for secrecy leakage for all functions, as considered in \cite{LifengFCTrans,ourISITSecurePrivateFunctArxiv}. Furthermore, we impose two privacy leakage constraints to minimize the information leakage about $X^n$ to the fusion center and eavesdropper because the same remote source would be measured if another function would be computed in the same network (see also \cite{bizimMMMMTIFS} for motivations to consider privacy leakage with respect to a remote source) as well as public storage constraints that minimize the rate of storage for transmitting nodes. 

We next define lossless and lossy single-function computation rate regions.

\subsection{Lossless Single-Function Computation}
Consider the single-function computation model  illustrated in Figure~\ref{fig:TIFSTwoTransmitHiddenFunction}. The corresponding lossless rate region is defined as follows.

\begin{Definition}\label{def:systemmodellossless}
	\normalfont A \emph{lossless} tuple $(R_{\text{s}}, R_{\text{w},1},R_{\text{w},2},R_{\ell,{\text{Dec}}}, R_{\ell,{\text{Eve}}})$ is \emph{achievable} if, for any $\delta\!>\!0$, there exist $n\!\geq\!1$, two encoders, and one decoder such that
	\begin{align}
	&\Pr\Big[f^n(\widetilde{X}_1^n,\widetilde{X}_2^n,Y^n) \neq \widehat{f^n}\Big] \leq \delta&&\!\!\!\!\! (\text{reliability})\label{eq:reliability_cons}\\
	&\frac{1}{n} I(\widetilde{X}^n_1,\widetilde{X}^n_2,Y^n;W_1,W_2|Z^n) \leq R_{\text{s}}+\delta&&\!\!\!\!\!(\text{secrecy})\label{eq:secrecyleakage_cons}\\
	&\frac{1}{n}\log\big|\mathcal{W}_1\big| \leq R_{\text{w,1}}+\delta&&\!\!\!\!\!(\text{storage 1})\label{eq:storage_cons1}\\
	&\frac{1}{n}\log\big|\mathcal{W}_2\big| \leq R_{\text{w,2}}+\delta&&\!\!\!\!\!(\text{storage 2})\label{eq:storage_cons2}\\
	&\frac{1}{n}I(X^n;W_1,W_2|Y^n) \leq R_{\ell,\text{Dec}}+\delta&&\!\!\!\!\!(\text{privacyDec})\label{eq:privDec_cons}\\
	&\frac{1}{n}I(X^n;W_1,W_2|Z^n) \leq R_{\ell,\text{Eve}}+\delta&&\!\!\!\!\!(\text{privacyEve})\label{eq:privEve_cons}.
	\end{align}
	The \emph{lossless} region $\mathcal{R}$ is the closure of the set of all achievable lossless tuples.\hfill $\lozenge$
\end{Definition}

\subsection{Lossy Single-Function Computation}
The corresponding lossy rate region for the single-function computation model  illustrated in Figure~\ref{fig:TIFSTwoTransmitHiddenFunction} is defined as follows.

\begin{Definition}\label{def:systemmodellossy}
	\normalfont A \emph{lossy} tuple $(R_{\text{s}}, R_{\text{w},1},R_{\text{w},2},R_{\ell,{\text{Dec}}}, R_{\ell,{\text{Eve}}},D)$ is \emph{achievable} if, for any $\delta\!>\!0$, there exist $n\!\geq\!1$, two encoders, and one decoder such that (\ref{eq:secrecyleakage_cons})-(\ref{eq:privEve_cons}) and
	\begin{align}
	& \mathbb{E}\Big[d(f^n(\widetilde{X}_1^n,\widetilde{X}_2^n,Y^n),\widehat{f^n})\Big] \leq D+\delta&&\!\!\!\!\!(\text{distortion})\label{eq:distortion_cons}
	\end{align}
	where
	\begin{align}
	d(f^n,\widehat{f^n})=\frac{1}{n}\sum_{i=1}^nd(f_i,\widehat{f}_i)\label{eq:disperletter}
	\end{align}
	is a per-letter distortion metric. The \emph{lossy} region $\mathcal{R}_{\text{D}}$ is the closure of the set of all achievable lossy tuples.\hfill $\lozenge$
\end{Definition}

\section{Inner and Outer Bounds}\label{sec:innerouterboundsforgeneral}

\subsection{Lossless Single-Function Computation}
We first extend the notion of \textit{admissibility} defined in \cite{CodingforComputing} for a single auxiliary random variable to two auxiliary random variables, used in the inner and outer bounds given below for lossless function computation; see also \cite[Theorem~3]{LifengFCTrans}.

\begin{Definition}\label{def:admissible}
	\normalfont A pair of (vector) random variables $(U_1,U_2)$ is admissible for a function $f(\widetilde{X}_1,\widetilde{X}_2,Y)$ if  we have
	\begin{align}
	H(f(\widetilde{X}_1,\widetilde{X}_2,Y)|\, U_1,U_2,Y) = 0
	\end{align}	
	and 
	\begin{align}
	&U_1-\widetilde{X}_1-(\widetilde{X}_2,Y)\\
	&U_2-\widetilde{X}_2-(\widetilde{X}_1,Y)
	\end{align}
	form Markov chains.\hfill $\lozenge$
\end{Definition}

We next provide inner and outer bounds for the lossless region $\mathcal{R}$; see Section~\ref{sec:proofinnerouter2lossless} for a proof sketch.

\begin{Theorem}\label{theo:innerouterfor2nodeslossless}
	\emph{(Inner Bound):} An achievable lossless region is the union over all $P_Q$, $P_{V_1|Q}$, $P_{V_2|Q}$, $P_{U_1|V_1}$, $P_{U_2|V_2}$, $P_{\widetilde{X}_1|U_1}$, and $P_{\widetilde{X}_2|U_2}$ of the rate tuples $(R_{\text{s}}, R_{\text{w},1},R_{\text{w},2},R_{\ell,{\text{Dec}}}, R_{\ell,{\text{Eve}}})$ such that $(U_1,U_2)$ pair is admissible for the function $f(\widetilde{X}_1,\widetilde{X}_2,Y)$ and
	\begin{align}
	&R_{\text{s}}\geq\Big[I(U_{1},U_2;Z|V_1,V_2,Q)-I(U_1,U_2;Y|V_1,V_2,Q)\Big]^- +I(U_1,U_2;\widetilde{X}_1,\widetilde{X}_2|Z)\label{eq:achRs}\\
	&R_{\text{w},1}\geq I(V_1;\widetilde{X}_1|V_2,Y)+I(U_1;\widetilde{X}_1|V_1,U_2,Y)\label{eq:achstorage1}\\
	&R_{\text{w},2}\geq I(V_2;\widetilde{X}_2|V_1,Y)+I(U_2;\widetilde{X}_2|U_1,V_2,Y)\label{eq:achstorage2}\\
	&R_{\text{w},1}+R_{\text{w},2}\geq I(U_2;\widetilde{X}_2|U_1,V_2,Y)+I(U_1;\widetilde{X}_1|V_1,V_2,Y)\nonumber\\
	&\qquad\qquad\quad\quad + I(V_2;\widetilde{X}_2|V_1,Y)+I(V_1;\widetilde{X}_1|Y)\label{eq:achsumstorage}\\
	&R_{\ell,\text{Dec}}\geq I(U_1,U_2;X|Y)\label{eq:achRellDec}\\
	&R_{\ell,\text{Eve}}\geq\Big[I(U_1,U_2;Z|V_1,V_2,Q)-I(U_1,U_2;Y|V_1,V_2,Q)\Big]^-+I(U_1,U_2;X|Z)	\label{eq:achRellEve}
	\end{align}
	where we have
	\begin{align}
	&P_{QV_1V_2U_1U_2\widetilde{X}_1\widetilde{X}_2XYZ}=P_{Q|V_1V_2}P_{V_1|U_1}P_{U_1|\widetilde{X}_1}P_{\widetilde{X}_1|X}P_{V_2|U_2}P_{U_2|\widetilde{X}_2}P_{\widetilde{X}_2|X}P_XP_{YZ|X}\label{eq:iid}.
	\end{align}
	
	\emph{(Outer Bound):} An outer bound for the lossless region $\mathcal{R}$ is the union of the rate tuples in (\ref{eq:achRs}), (\ref{eq:achsumstorage})-(\ref{eq:achRellEve}), and
	\begin{align}
	&R_{\text{w},1}\geq I(V_{1};\widetilde{X}_{1}|V_{2},Y)+I(U_{1};\widetilde{X}_{1}|V_{1},U_{2},Y)\nonumber\\
	&\qquad\qquad -I(V_{1};V_{2}|\widetilde{X}_{1},Y) - I(U_{1};U_{2}|\widetilde{X}_{1},Y,V_{1})\label{eq:conversestorage1}\\
	&R_{\text{w},2}\geq I(V_{2};\widetilde{X}_{2}|V_{1},Y)+I(U_{2};\widetilde{X}_{2}|U_{1},V_{2},Y)\nonumber\\
	&\qquad\qquad -I(V_{2};V_{1}|\widetilde{X}_{2},Y)- I(U_{2};U_{1}|\widetilde{X}_{2},Y,V_{2})\label{eq:conversestorage2}
	\end{align}
	over all $P_Q$, $P_{V_1|Q}$, $P_{V_2|Q}$, $P_{U_1|V_1}$, $P_{U_2|V_2}$, $P_{\widetilde{X}_1|U_1}$, and $P_{\widetilde{X}_2|U_2}$ such that $(U_1,U_2)$ pair is admissible for the function $f(\widetilde{X}_1,\widetilde{X}_2,Y)$ and 
	\begin{align}
	&(Q,V_1)-U_1-\widetilde{X}_{1}-X-(\widetilde{X}_2,Y,Z)\label{eq:Markovconverse1}\\
	&(Q,V_2)-U_2-\widetilde{X}_{2}-X-(\widetilde{X}_1,Y,Z)\label{eq:Markovconverse2}
	\end{align}
	form Markov chains. One can limit the cardinalities to $|\mathcal{Q}|~\leq~ 2$, $|\mathcal{V}_1|\leq |\widetilde{X}_1|+6$, $|\mathcal{V}_2|\leq |\widetilde{X}_2|+6$, $|\mathcal{U}_1|\leq (|\widetilde{X}_1|+6)^2$, and $|\mathcal{U}_2|\leq (|\widetilde{X}_2|+6)^2$.
\end{Theorem}

We remark that if the joint probability distribution in (\ref{eq:iid}) is imposed on the outer bound, (\ref{eq:conversestorage1}) and (\ref{eq:conversestorage2}) recover (\ref{eq:achstorage1}) and (\ref{eq:achstorage2}), respectively, because then 
\begin{align}
&(V_1,U_1)-\widetilde{X}_1-(Y,U_2,V_2)\\ &(V_2,U_2)-\widetilde{X}_2-(Y,U_1,V_1)
\end{align}
form Markov chains for (\ref{eq:iid}). However, the outer bound that satisfies (\ref{eq:Markovconverse1}) and (\ref{eq:Markovconverse2}) defines a rate region that is in general larger than the rate region defined by the inner bound that satisfies (\ref{eq:iid}). Thus, inner and outer bounds generally differ. The results in Theorem~\ref{theo:innerouterfor2nodeslossless} recovers previous results including \cite[Theorem~3]{LifengFCTrans} and, naturally, also other results that are recovered by these previous results such as the SW coding region.

\subsection{Lossy Single-Function Computation}
We next provide inner and outer bounds for the lossy region $\mathcal{R}_{\text{D}}$; see below for a proof sketch.

\begin{Theorem}\label{theo:innerouterlossyfor2nodes}
	\emph{(Inner Bound):} An achievable lossy region is the union over all $P_Q$, $P_{V_1|Q}$, $P_{V_2|Q}$, $P_{U_1|V_1}$, $P_{U_2|V_2}$, $P_{\widetilde{X}_1|U_1}$, and $P_{\widetilde{X}_2|U_2}$ of the rate tuples in (\ref{eq:achRs})-(\ref{eq:achRellEve}) and
	\begin{align}
	&D\geq \mathbb{E}[d(f(\widetilde{X}_1,\widetilde{X}_2,Y),g(U_1,U_2,Y))]\label{eq:distortioninnerlossy}
	\end{align}
	for some function $g(\cdot,\cdot,\cdot)$ and where $P_{QV_1V_2U_1U_2\widetilde{X}_1\widetilde{X}_2XYZ}$ is equal to (\ref{eq:iid}).
	
	\emph{(Outer Bound):} An outer bound for the lossy region $\mathcal{R}_{\text{D}}$ is the union over all $P_Q$, $P_{V_1|Q}$, $P_{V_2|Q}$, $P_{U_1|V_1}$, $P_{U_2|V_2}$, $P_{\widetilde{X}_1|U_1}$, and $P_{\widetilde{X}_2|U_2}$ of the set of rate tuples $(R_{\text{s}}, R_{\text{w},1},R_{\text{w},2},R_{\ell,{\text{Dec}}}, R_{\ell,{\text{Eve}}},D)$ in (\ref{eq:achRs}), (\ref{eq:achsumstorage})-(\ref{eq:achRellEve}), (\ref{eq:conversestorage1}), (\ref{eq:conversestorage2}), and (\ref{eq:distortioninnerlossy}) such that (\ref{eq:Markovconverse1}) and (\ref{eq:Markovconverse2}) form Markov chains. One can limit the cardinalities to $|\mathcal{Q}|~\leq~2$, $|\mathcal{V}_1|\leq |\widetilde{X}_1|+7$, $|\mathcal{V}_2|\leq |\widetilde{X}_2|+7$, $|\mathcal{U}_1|\leq (|\widetilde{X}_1|+7)^2$, and $|\mathcal{U}_2|\leq (|\widetilde{X}_2|+7)^2$.
	
\end{Theorem}

\begin{proof}[Proof Sketch] 
	The achievability proof of the lossy function computation problem follows from the achievability proof of its lossless version given in Section~\ref{subsec:innerboundproof2lossless} by replacing the admissibility constraint with the constraint that $P_{U_1|\widetilde{X}_1}$, $P_{V_1|U_1}$, $P_{U_2|\widetilde{X}_2}$, and $P_{V_2|U_2}$ are chosen such that there exists a function $g(U_1,U_2,Y)$ that satisfies
	\begin{align}
	&g^n(U_1^n,U_2^n,Y^n)=\{g(U_{1,i},U_{2,i},Y_{i})\}_{i=1}^n\\
	&\mathbb{E}[d(f^n(\widetilde{X}_1^n,\widetilde{X}_2^n,Y^n),g^n(U_1^n,U^n_2,Y^n))]\leq D+\epsilon_n
	\end{align}
	where $\epsilon_n>0$ such that $\epsilon_n\rightarrow 0$ when $n\rightarrow\infty$. Since all $(\widetilde{x}_1^n,\widetilde{x}_2^n,y^n,u_1^n,u_2^n)$ tuples are in the jointly typical set with high probability, by the typical average lemma \cite[pp.~26]{Elgamalbook}, constraint in (\ref{eq:distortion_cons}) is satisfied. 
	
	The proof of the outer bound applies the standard properties of the Shannon entropy and follows mainly from the outer bound proof for the lossless function computation problem given in Section~\ref{subsec:outerboundproof2lossless}. However, the proof for the lossless function computation problem requires the auxiliary random variables to be admissible as defined in Definition~\ref{def:admissible}, unlike the lossy function computation problem. Thus, the outer bound proof for Theorem~\ref{theo:innerouterlossyfor2nodes} follows by replacing the admissibility step (\ref{eq:admissibilityandFano}) in the outer bound proof for the lossless function computation problem with the step
	\begin{align}
	&n(D+\delta_n)\nonumber\\
	&\overset{(a)}{\geq}\mathbb{E}\Big[\sum_{i=1}^nd\left(f_i(\widetilde{X}_{1,i},\widetilde{X}_{2,i},Y_i),\widehat{f_i}(W_1,W_2,Y^n)\right)\Big]\nonumber\\
	&\overset{(b)}{\geq}\!\mathbb{E}\!\Big[\!\sum_{i=1}^n\!d\left(\!f_i(\widetilde{X}_{1,i},\widetilde{X}_{2,i},Y_i),g_i(W_1,W_2,Y^n,X^{i-1}\!,Z^{i-1})\!\right)\!\Big]\nonumber\\
	&\overset{(c)}{=}\!\mathbb{E}\!\Big[\!\sum_{i=1}^n\!d\left(\!f_i(\widetilde{X}_{1,i},\widetilde{X}_{2,i},Y_i),g_i(W_1,W_2,Y_{i}^n, X^{i-1}\!,Z^{i-1})\!\right)\!\Big]\nonumber\\
	&\overset{(d)}{=}\mathbb{E}\Big[\sum_{i=1}^nd\left(f(\widetilde{X}_{1,i},\widetilde{X}_{2,i},Y_i),g(U_{1,i},U_{2,i},Y_i)\right)\Big]\label{eq:converseforsinglelossy}
	\end{align}
	where $(a)$ follows by (\ref{eq:distortion_cons}) and (\ref{eq:disperletter}), $(b)$ follows since there exists a function $g_i(\cdot,\cdot,\cdot)$ that achieves a distortion that is not greater than the distortion achieved by $\widehat{f_i}(W_1,W_2,Y^n)$, where the distortion is measured with respect to $f_i(\widetilde{X}_{1,i},\widetilde{X}_{2,i},Y_i)$, since $g_i(\cdot,\cdot,\cdot)$ has additional inputs, $(c)$ follows from the Markov chain given in (\ref{eq:conv_Markov2}), and $(d)$ follows from the definitions of $U_{1,i}$ and $U_{2,i}$ given in (\ref{eq:defU1}) and (\ref{eq:defU2}), respectively. Furthermore, the proof of the cardinality bounds for the lossy case follows from the proof for the lossless case since we preserve the same probability and conditional entropy values as being preserved for the lossless function computation problem with the addition of preserving the value of $g(U_1,U_2,Y)=g(U_1,U_2,V_1,V_2,Y)$, following from the Markov chain
	\begin{align}
	(V_1,V_2)-(U_1,U_2,Y)-g(U_1,U_2,Y).  
	\end{align}
\end{proof}

Entirely similar to Theorem~\ref{theo:innerouterfor2nodeslossless}, the inner and outer bounds given in Theorem~\ref{theo:innerouterlossyfor2nodes} do not match in general because of different Markov chains imposed.

\begin{Remark}
	\normalfont Since all secrecy and privacy rate terms given in the outer bounds in Theorems~\ref{theo:innerouterfor2nodeslossless} and ~\ref{theo:innerouterlossyfor2nodes}, i.e., lower bounds in (\ref{eq:achRs}), (\ref{eq:achRellDec}), and (\ref{eq:achRellEve}), are generally strictly positive, strong secrecy or strong privacy constraints cannot be satisfied in general for the lossless and lossy single-function computation problems.
\end{Remark}

We next provide the simplified rate region bounds for various sets of computed functions $f(\cdot,\cdot,\cdot)$ and measurement channels~$P_{YZ|X}$.

\section{Rate Regions for Special Sets of Computed Functions and Measurement Channels}\label{sec:specialcases}
The terms that characterize the rate region bounds for the lossless and lossy function computation problems for various sets of functions and channels are the same, except (1) removal of the admissibility requirement; (2) addition of a distortion constraint; and (3) increase in the cardinality bounds on the auxiliary random variables for the lossy case as compared to the lossless case. Thus, we provide simplified rate region bounds only for the lossless case. However, we remark that the optimal auxiliary random variables for lossless and lossy cases might differ. Therefore, the corresponding lossless and lossy rate regions might look different for the same joint probability distribution $P_{\widetilde{X}_1\widetilde{X}_2XYZ}$.
\subsection{Partially-Invertible Functions}
We now impose the condition that the function $f(\widetilde{X}_1,\widetilde{X}_2,Y)$ is \emph{partially-invertible} with respect to $\widetilde{X}_1$, i.e., we have \cite{successivesourcecoingEricsonKorner,TchamkertenCompute}
\begin{align}
H(\widetilde{X}_1|f(\widetilde{X}_1,\widetilde{X}_2,Y),Y)=0.
\end{align}
For such functions, it is straightforward to show that we have the following achievable rate region for the lossless function computation problem with two transmitting nodes. We remark that the proof of Lemma~\ref{lem:partialinvertible} follows from the inner bound in Theorem~\ref{theo:innerouterfor2nodeslossless} by assigning $U_1=\widetilde{X}_1$ and the corresponding outer bound can be similarly obtained from Theorem~\ref{theo:innerouterfor2nodeslossless}. Furthermore, by symmetry the lossy rate region bounds for a function $f(\widetilde{X}_1,\widetilde{X}_2,Y)$ that is partially invertible with respect to $\widetilde{X}_2$ can be obtained by assigning $U_2=\widetilde{X}_2$.

\begin{Lemma}\label{lem:partialinvertible}
	The lossless region $\mathcal{R}$ when $f(\widetilde{X}_1,\widetilde{X}_2,Y)$ is a partially invertible function with respect to $\widetilde{X}_1$ includes the set of all tuples $(R_{\text{s}}, R_{\text{w},1},R_{\text{w},2},R_{\ell,{\text{Dec}}}, R_{\ell,{\text{Eve}}})$ such that $U_2$ is admissible for the function $f(\widetilde{X}_1,\widetilde{X}_2,Y)$ and 
	\begin{align}
		&R_{\text{s}}\geq \Big[I(\widetilde{X}_{1},U_2;Z|V_1,V_2,Q)-I(\widetilde{X}_1,U_2;Y|V_1,V_2,Q)\Big]^-\!+\!H(\widetilde{X}_1|Z)\!+\!I(U_2;\widetilde{X}_2|\widetilde{X}_1,Z)\\
		&R_{\text{w},1}\geq H(\widetilde{X}_1|V_2,Y)-I(\widetilde{X}_1;U_2|V_1,V_2,Y)\nonumber\\
		&R_{\text{w},2}\geq I(V_2;\widetilde{X}_2|V_1,Y)+I(U_2;\widetilde{X}_2|\widetilde{X}_1,V_2,Y)\\
		&R_{\text{w},1}+R_{\text{w},2}\geq I(U_2;\widetilde{X}_2|\widetilde{X}_1,V_2,Y)+H(\widetilde{X}_1|V_1,V_2,Y) + I(V_2;\widetilde{X}_2|V_1,Y)+I(V_1;\widetilde{X}_1|Y)\\
		&R_{\ell,\text{Dec}}\geq I(\widetilde{X}_1,U_2;X|Y)\\
		&R_{\ell,\text{Eve}}\geq\Big[I(\widetilde{X}_1,U_2;Z|V_1,V_2,Q)-I(\widetilde{X}_1,U_2;Y|V_1,V_2,Q)\Big]^-+I(\widetilde{X}_1, U_2;X|Z)
	\end{align}
	for some function $\ell(\cdot,\cdot,\cdot)$ such that (\ref {eq:iid}) follows with $U_1=\widetilde{X}_1$.
\end{Lemma}

\subsection{Invertible Functions}
Suppose now we impose the condition that the function $f(\widetilde{X}_1,\widetilde{X}_2,Y)$ is \emph{invertible}, i.e., we have \cite{successivesourcecoingEricsonKorner,TchamkertenCompute}
\begin{align}
H(\widetilde{X}_1,\widetilde{X}_2|f(\widetilde{X}_1,\widetilde{X}_2,Y),Y)=0.
\end{align}
We provide in Lemma~\ref{lem:invertible} below an achievable rate region for the lossless computation problem with two transmitting nodes when the function $f(\widetilde{X}_1,\widetilde{X}_2,Y)$ is invertible. The proof of Lemma~\ref{lem:invertible} follows from Theorem~\ref{theo:innerouterfor2nodeslossless} by assigning $U_1=\widetilde{X}_1$, $U_2=\widetilde{X}_2$, and constant $V_1$ and $V_2$. Note that choosing $V_1$ and $V_2$ constant results generally in suboptimal rate regions.

\begin{Lemma}\label{lem:invertible}
	The lossless rate region $\mathcal{R}$ when $f(\widetilde{X}_1,\widetilde{X}_2,Y)$ is an invertible function includes the set of all tuples $(R_{\text{s}}, R_{\text{w},1},R_{\text{w},2},R_{\ell,{\text{Dec}}}, R_{\ell,{\text{Eve}}})$ satisfying
	\begin{align}
	&R_{\text{s}}\geq\big[I(\widetilde{X}_{1},\widetilde{X}_2;Z|Q)-I(\widetilde{X}_1,\widetilde{X}_2;Y|Q)\big]^-+H(\widetilde{X}_1,\widetilde{X}_2|Z)\label{eq:Rscor}\\
	&R_{\text{w},1}\geq H(\widetilde{X}_1|\widetilde{X}_2,Y)\label{eq:storage1cor}\\
	&R_{\text{w},2}\geq H(\widetilde{X}_2|\widetilde{X}_1,Y)\label{eq:storage2cor}\\
	&R_{\text{w},1}+R_{\text{w},2}\geq H(\widetilde{X}_1,\widetilde{X}_2|Y)\label{eq:sumstoragecor}\\
	&R_{\ell,\text{Dec}}\geq I(\widetilde{X}_1,\widetilde{X}_2;X|Y)\label{eq:RellDeccor}\\
	&R_{\ell,\text{Eve}}\geq\big[I(\widetilde{X}_1,\widetilde{X}_2;Z|Q)-I(\widetilde{X}_1,\widetilde{X}_2;Y|Q)\big]^- +I(\widetilde{X}_1,\widetilde{X}_2;X|Z)	\label{eq:RellEvecor}
	\end{align}
	where $Q-(\widetilde{X}_1,\widetilde{X}_2)-X-(Y,Z)$ form a Markov chain. One can limit the cardinality to $|\mathcal{Q}|\leq 2$.
\end{Lemma}

\subsection{Invertible Functions and Two Different Degraded Channels}
The lossless rate region given in Lemma~\ref{lem:invertible} can be further simplified by imposing conditions on the measurement channel $P_{YZ|X}$ in addition to the function $f(\widetilde{X}_1,\widetilde{X}_2,Y)$ being invertible. We next establish achievable lossless rate regions for two different physically degraded channels.

\subsubsection{Eve's Channel is Physically-Degraded}
Suppose the measurement channel $P_{YZ|X}$ is physically-degraded such that
\begin{align}
P_{YZ|X} = P_{Y|X}P_{Z|Y}.\label{eq:phydegraded}
\end{align}
For invertible functions and physically degraded measurement channels $P_{YZ|X}$ as defined in (\ref{eq:phydegraded}), we provide an achievable lossless rate region in Lemma~\ref{lem:invertibledegradedlossless}. The proof of Lemma~\ref{lem:invertibledegradedlossless} follows from Lemma~\ref{lem:invertible} and by using the following Markov chain for this case 
\begin{align}
&(\widetilde{X}_1,\widetilde{X}_2)-X-Y-Z
\end{align}
which follows by (\ref{eq:phydegraded}).

\begin{Lemma}\label{lem:invertibledegradedlossless}
	The lossless rate region $\mathcal{R}$ when $f(\widetilde{X}_1,\widetilde{X}_2,Y)$ is an invertible function and $P_{YZ|X}$ is as given in (\ref{eq:phydegraded}) includes the set of all tuples $(R_{\text{s}}, R_{\text{w},1},R_{\text{w},2},R_{\ell,{\text{Dec}}}, R_{\ell,{\text{Eve}}})$ satisfying (\ref{eq:storage1cor})--(\ref{eq:RellDeccor}) and 
	\begin{align}
	&R_{\text{s}}\geq H(\widetilde{X}_1,\widetilde{X}_2|Y)\label{eq:Rscorinvdegr}\\
	&R_{\ell,\text{Eve}}\geq I(\widetilde{X}_1,\widetilde{X}_2;X|Y)	\label{eq:RellEvecordegradedinv}.
	\end{align}
\end{Lemma}

\subsubsection{Fusion Center's Channel is Physically-Degraded}
Suppose the measurement channel $P_{YZ|X}$ is physically-degraded such that
\begin{align}
P_{YZ|X} = P_{Z|X}P_{Y|Z}.\label{eq:phydegradedcase2}
\end{align}
For invertible functions and physically degraded measurement channels $P_{YZ|X}$ as defined in (\ref{eq:phydegradedcase2}), we provide an achievable lossless rate region in Lemma~\ref{lem:invertibledegradedlosslesscase2}. The proof of Lemma~\ref{lem:invertibledegradedlosslesscase2} follows from Lemma~\ref{lem:invertible} and by using the following Markov chain for this case 
\begin{align}
&(\widetilde{X}_1,\widetilde{X}_2)-X-Z-Y
\end{align}
which follows by (\ref{eq:phydegradedcase2}).

\begin{Lemma}\label{lem:invertibledegradedlosslesscase2}
	The lossless rate region $\mathcal{R}$ when $f(\widetilde{X}_1,\widetilde{X}_2,Y)$ is an invertible function and $P_{YZ|X}$ is as given in (\ref{eq:phydegradedcase2}) includes the set of all tuples $(R_{\text{s}}, R_{\text{w},1},R_{\text{w},2},R_{\ell,{\text{Dec}}}, R_{\ell,{\text{Eve}}})$ satisfying (\ref{eq:storage1cor})--(\ref{eq:RellDeccor}) and 
	\begin{align}
	&R_{\text{s}}\geq H(\widetilde{X}_1,\widetilde{X}_2|Z)\label{eq:Rscorinvdegrcase2}\\
	&R_{\ell,\text{Eve}}\geq I(\widetilde{X}_1,\widetilde{X}_2;X|Z)	\label{eq:RellEvecordegradedinvcase2}.
	\end{align}
\end{Lemma}

\begin{Remark}
	\normalfont The rate regions given in Lemmas~\ref{lem:invertible}-\ref{lem:invertibledegradedlosslesscase2} can be plotted by computing the terms that characterize the regions since $P_{\widetilde{X}_1\widetilde{X}_2XYZ}$ is fixed for function computation problems considered. However, the rate region given in Lemma~\ref{lem:partialinvertible}, similar to the inner bounds given in Theorems~\ref{theo:innerouterfor2nodeslossless} and \ref{theo:innerouterlossyfor2nodes}, might not be easy to characterize due to the requirement to optimize the auxiliary random variables whose cardinalities are bounded by large terms. Thus, evaluating the rate region for a function computation problem with two transmitting terminals is generally significantly more difficult than characterization of the rate region for function computation with one transmitting terminal; see \cite{ourISITSecurePrivateFunctArxiv} for an information bottleneck example for the latter problem. 
\end{Remark}

We next evaluate an achievable lossless rate region $\mathcal{R}$ by using Lemma~\ref{lem:invertibledegradedlosslesscase2} for specific measurement channels when $f(\widetilde{X}_1,\widetilde{X}_2,Y)$ is an invertible function.

\subsection{Lossless Rate Region Example}

Suppose measurement channels in Figure~\ref{fig:TIFSTwoTransmitHiddenFunction} have binary input and output alphabets with multiplicative Bernoulli noise components, i.e., we have $\mathcal{X}=\mathcal{\widetilde{X}}_1=\mathcal{\widetilde{X}}_2=\mathcal{Z}=\mathcal{Y}=\mathcal{S}_1=\mathcal{S}_2=\mathcal{S}_Z=\mathcal{S}_Y=\{0,1\}$ and
\begin{align}
\widetilde{X}_1 =S_1\cdot X,\qquad\qquad \widetilde{X}_2=S_2\cdot X,\qquad\qquad Z=S_Z\cdot X, \qquad\qquad Y=S_Y\cdot X
\end{align}
where $S_1$, $S_2$, $X$, and $(S_Z,S_Y)$ are mutually independent, and we have $P_X(1)=0.5$, $P_{S_1}(1)=\beta_1$, $P_{S_2}(1)=\beta_2$, $P_{S_ZS_Y}(0,0)=(1\!-\!q)$, $P_{S_ZS_Y}(1,1)=q\alpha$, and $P_{S_ZS_Y}(1,0)=q(1\!-\!\alpha)$ for fixed $\beta_1,\beta_2, q,\alpha\in[0,1]$, so (\ref{eq:phydegradedcase2}) is satisfied; see also \cite[Section~IV-A]{MariMicheleBCJCAS}. Using Lemma~\ref{lem:invertibledegradedlosslesscase2} for the given probability distributions, we evaluate an achievable lossless rate region $\mathcal{R}$ for an invertible function computation scenario with two transmitting nodes, in which, e.g., $\beta_1 = 0.2$, $\beta_2=0.11$, $\alpha = 0.3$, and $q=0.25$ and obtain a lossless rate region that is characterized by
\begin{alignat}{2}
	&R_{\text{s}}\geq 0.7579\; \text{bits/symbol},\qquad &&R_{\text{w},1}\geq 0.4626\; \text{bits/symbol},\\
	& R_{\text{w},2}\geq 0.3021\; \text{bits/symbol},\qquad &&R_{\text{w},1}+R_{\text{w},2}\geq 0.7686\; \text{bits/symbol},\\
	&R_{\ell,\text{Dec}}\geq 0.1577\; \text{bits/symbol},\qquad&& R_{\ell,\text{Eve}}\geq 0.1469\;\text{bits/symbol}
\end{alignat}
where the sum-storage rate constraint is active since the sum of the bounds on $R_{\text{w},1}$ and $R_{\text{w},2}$ is smaller than the bound on $(R_{\text{w},1}+R_{\text{w},2})$.

\section{Proof of Theorem~\ref{theo:innerouterfor2nodeslossless}}\label{sec:proofinnerouter2lossless}
\subsection{Inner Bound}\label{subsec:innerboundproof2lossless}
\begin{proof}[Proof Sketch]
	The OSRB method \cite{AminOSRB} is used for the proof of achievability by applying the steps given in \cite[Section~1.6]{BlochLectureNotes2018}. Let
	\begin{align}
	(V_1^n,V_2^n,U_1^n,U_2^n,\widetilde{X}_1^n,\widetilde{X}_2^n,X^n,Y^n,Z^n)\label{eq:nletterrvslateriid}
	\end{align}	
	be i.i.d. according to $P_{V_1V_2U_1U_2\widetilde{X}_1\widetilde{X}_2XYZ}$ that can be obtained from (\ref{eq:iid}) with fixed $P_{U_1|\widetilde{X}_1}$, $P_{V_1|U_1}$, $P_{U_2|\widetilde{X}_2}$, and $P_{V_2|U_2}$ such that the pair $(U_1,U_2)$ is admissible for a function $f(\widetilde{X}_1,\widetilde{X}_2,Y)$, so $(U_1^n,U_2^n)$ is also admissible since random variables in (\ref{eq:nletterrvslateriid}) are i.i.d.
	
	To each $v_1^n$ assign two random bin indices $(F_{\text{v}_1},W_{\text{v}_1})$ such that 	$F_{\text{v}_1}\in[1:2^{n\widetilde{R}_{\text{v}_1}}]$ and $W_{\text{v}_1}\in[1:2^{nR_{\text{v}_1}}]$. Furthermore, to each $u_1^n$ assign two random indices $(F_{\text{u}_1},W_{\text{u}_1})$ such that $F_{\text{u}_1}\in[1:2^{n\widetilde{R}_{\text{u}_1}}]$ and $W_{\text{u}_1}\in[1:2^{nR_{\text{u}_1}}]$. Similarly, random indices $(F_{\text{v}_2},W_{\text{v}_2})$ and  $(F_{\text{u}_2},W_{\text{u}_2})$ are assigned to each $v_2^n$ and $u_2^n$, respectively. The indices $F_1=(F_{\text{v}_1}, F_{\text{u}_1})$, and $F_2=(F_{\text{v}_2},F_{\text{u}_2})$ represent the public choice of two encoders and one decoder, whereas $W_1=(W_{\text{v}_1}, W_{\text{u}_1})$ and $W_2=(W_{\text{v}_2}, W_{\text{u}_2})$ are the public messages sent by the encoders $\Enc_1(\cdot)$ and $\Enc_2(\cdot)$, respectively, to the fusion center.
	
	We consider the following decoding order: 
	\begin{enumerate}
		\item observing $(Y^n,F_{\text{v}_1},W_{\text{v}_1})$, the decoder $\Dec(\cdot)$ estimates $V_1^n$ as $\widehat{V}_1^n$;\label{decorder:Step1}
		\item observing $(Y^n,\widehat{V}_1^n,F_{\text{v}_2},W_{\text{v}_2})$, the decoder estimates $V^n_2$ as $\widehat{V}_2^n$;\label{decorder:Step2}
		\item observing $(Y^n,\widehat{V}_1^n,\widehat{V}_2^n,F_{\text{u}_1},W_{\text{u}_1})$, the decoder estimates $U^n_1$ as $\widehat{U}_1^n$;\label{decorder:Step3}
		\item observing $(Y^n,\widehat{V}_1^n,\widehat{V}_2^n,\widehat{U}_1^n,F_{\text{u}_2},W_{\text{u}_2})$, the decoder estimates $U^n_2$ as $\widehat{U}_2^n$.\label{decorder:Step4}
	\end{enumerate}
	By swapping indices $1$ and $2$ in the decoding order another corner point in the achievable rate region is obtained, so we analyze the given decoding order but also provide the results for the other corner point.
	
	Consider Step~\ref{decorder:Step1} in the decoding order given above. Using a SW \cite{SW} decoder, one can reliably estimate $V_1^n$ from $(Y^n,F_{\text{v}_1},W_{\text{v}_1})$ such that the expected value of the error probability taken over the random bin assignments vanishes when $n\rightarrow\infty$, if we have \cite[Lemma 1]{AminOSRB}
	\begin{align}
	\widetilde{R}_{\text{v}_1} + R_{\text{v}_1}> H(V_1|Y).\label{eq:Step1reconstr}
	\end{align}	
	Similarly, Step~\ref{decorder:Step2}, \ref{decorder:Step3}, and \ref{decorder:Step4} estimations are reliable if we have
	\begin{align}
	&\widetilde{R}_{\text{v}_2} + R_{\text{v}_2}> H(V_2|V_1,Y)\label{eq:Step2reconstr}\\
	&\widetilde{R}_{\text{u}_1} + R_{\text{u}_1}> H(U_1|V_1,V_2,Y)\label{eq:Step3reconstr}\\
	&\widetilde{R}_{\text{u}_2} +R_{\text{u}_2}> H(U_2|V_1,V_2,U_1,Y)\!\overset{(a)}{=}\!H(U_2|V_2,U_1,Y)\label{eq:Step4reconstr}
	\end{align} 
	where $(a)$ follows from the Markov chain $V_1-U_1-(U_2,V_2,Y)$. Therefore, (\ref{eq:reliability_cons}) is satisfied if (\ref{eq:Step1reconstr})-(\ref{eq:Step4reconstr}) are satisfied.

	The public index $F_{\text{v}_1}$ is almost independent of $\widetilde{X}_1^n$, so it is almost independent of $(\widetilde{X}_1^n,\widetilde{X}_2^n,X^n,Y^n,Z^n)$, if we have \cite[Theorem 1]{AminOSRB}
	\begin{align}
	\widetilde{R}_{\text{v}_1}<H(V_1|\widetilde{X}_1)\label{eq:independenceofFv1}
	\end{align} 
	because then the expected value, which is taken over the random bin assignments, of the variational distance between the joint probability distributions $\text{Unif}[1\!\!:\!2^{n\widetilde{R}_{\text{v}_1}}]\cdot P_{\widetilde{X}_1^n}$ and $P_{F_{\text{v}_1}\widetilde{X}_1^n}$ vanishes when $n\rightarrow\infty$. Furthermore, the public index $F_{\text{u}_1}$ is almost independent of $(V_1^n,\widetilde{X}_1^n)$, so it is almost independent of $(V_1^n,\widetilde{X}_1^n,\widetilde{X}_2^n,X^n,Y^n,Z^n)$, if we have 
	\begin{align}
	\widetilde{R}_{\text{u}_1}<H(U_1|V_1,\widetilde{X}_1).\label{eq:independenceofFu1}
	\end{align} 
	Similarly, $F_{\text{v}_2}$ is almost independent of $\widetilde{X}_2^n$ if we have
	\begin{align}
	\widetilde{R}_{\text{v}_2}<H(V_2|\widetilde{X}_2)\label{eq:independenceofFv2}
	\end{align}
	and $F_{\text{u}_2}$ is almost independent of $(V_2^n,\widetilde{X}_2^n)$ if we have
	\begin{align}
	\widetilde{R}_{\text{u}_2}<H(U_2|V_2,\widetilde{X}_2).\label{eq:independenceofFu2}
	\end{align} 
	
	To satisfy (\ref{eq:Step1reconstr})-(\ref{eq:independenceofFu2}), for any $\epsilon>0$ we fix 
	\begin{align}
	&\widetilde{R}_{\text{v}_1} = H(V_1|\widetilde{X}_1)-\epsilon\label{eq:R_v1tildechosen}\\
	&R_{\text{v}_1} = I(V_1;\widetilde{X}_1)-I(V_1;Y)+2\epsilon\label{eq:R_v1chosen}\\
	&\widetilde{R}_{\text{v}_2} = H(V_2|\widetilde{X}_2)-\epsilon\label{eq:R_v2tildechosen}\\
	&R_{\text{v}_2} = I(V_2;\widetilde{X}_2)-I(V_2;V_1,Y)+2\epsilon\label{eq:R_v2chosen}\\
	&\widetilde{R}_{\text{u}_1} = H(U_1|V_1,\widetilde{X}_1)-\epsilon\label{eq:R_u1tildechosen}\\
	&R_{\text{u}_1} = I(U_1;\widetilde{X}_1|V_1)-I(U_1;V_2,Y|V_1)+2\epsilon\label{eq:R_u1chosen}\\
	&\widetilde{R}_{\text{u}_2} = H(U_2|V_2,\widetilde{X}_2)-\epsilon\label{eq:R_u2tildechosen}\\
	&R_{\text{u}_2} = I(U_2;\widetilde{X}_2|V_2)-I(U_2;U_1,Y|V_2)+2\epsilon\label{eq:R_u2chosen}.
	\end{align}
	
	\textbf{Public Message (Storage) Rates}: (\ref{eq:R_v1chosen}) and (\ref{eq:R_u1chosen}) result in a public message (storage) rate $R_{\text{w}_1}$ of
	\begin{align}
	&R_{\text{w}_1} = R_{\text{v}_1} + R_{\text{u}_1}\nonumber\\
	&\overset{(a)}{=}  I(V_1;\widetilde{X}_1|Y)+H(U_1|V_1,V_2,Y)-H(U_1|V_1,\widetilde{X}_1)+4\epsilon \nonumber\\
	&\overset{(b)}{=}   I(V_1;\widetilde{X}_1|Y)+I(U_1;\widetilde{X}_1|V_1,V_2,Y)+4\epsilon\label{eq:R_wchosen}
	\end{align}
	where $(a)$ follows because $V_1-\widetilde{X}_1-Y$ form a Markov chain and $(b)$ follows because $U_1-(V_1,\widetilde{X}_1)-(V_2,Y)$ form a Markov chain. Furthermore,  (\ref{eq:R_v2chosen}) and (\ref{eq:R_u2chosen}) result in a storage rate $R_{\text{w}_2}$ of
	\begin{align}
	&R_{\text{w}_2} = R_{\text{v}_2} + R_{\text{u}_2}\nonumber\\
	&\overset{(a)}{=}  I(V_2;\widetilde{X}_2|V_1,Y)+H(U_2|U_1,V_2,Y)-H(U_2|V_2,\widetilde{X}_2)+4\epsilon \nonumber\\
	&\overset{(b)}{=}  I(V_2;\widetilde{X}_2|V_1,Y)+I(U_2;\widetilde{X}_2|U_1,V_2,Y)+4\epsilon
	\end{align}
	where $(a)$ follows from the Markov chain $V_2-\widetilde{X}_2-(V_1,Y)$ and $(b)$ from $U_2-(V_2,\widetilde{X}_2)-(U_1,Y)$. We remark that if the indices 1 and 2 in the decoding order given above are swapped, the other corner point with
	\begin{align}
	&R_{\text{w}_1}' =  I(V_1;\widetilde{X}_1|V_2,Y)+I(U_1;\widetilde{X}_1|U_2,V_1,Y)+4\epsilon\\
	&R_{\text{w}_2}' =  I(V_2;\widetilde{X}_2|Y)+I(U_2;\widetilde{X}_2|V_1,V_2,Y)+4\epsilon
	\end{align}
	is achieved.
	
	\textbf{Privacy Leakage to Decoder}: We have
	\begin{align}
	&I(X^n;W_1,W_2,F_1,F_2|Y^n)\nonumber\\
	& = I(X^n;W_1,W_2|F_1,F_2,Y^n)+I(X^n;F_1,F_2|Y^n)\nonumber\\
	&\overset{(a)}{\leq} H(X^n|Y^n) - H(X^n|W_1,W_2,F_1,F_2,V_1^n, V_2^n,U_1^n,U_2^n,Y^n)+4\epsilon_n\nonumber\\
	&\overset{(b)}{=}H(X^n|Y^n) - H(X^n|U_1^n,U_2^n,Y^n)+4\epsilon_n\nonumber\\
	& \overset{(c)}{=}nI(U_1,U_2;X|Y)+4\epsilon_n\label{eq:ach_privtoDec}
	\end{align}
	where \\
	$(a)$ follows for some $\epsilon_n>0$ with $\epsilon_n\rightarrow 0$ when $n\rightarrow\infty$ because
	\begin{align}
	&I(X^n;F_1,F_2|Y^n) \nonumber\\
	&= I(X^n;F_{\text{v}_1}|Y^n)+I(X^n;F_{\text{u}_1}|F_{\text{v}_1},Y^n)+I(X^n;F_{\text{v}_2}|F_{\text{v}_1},F_{\text{u}_1},Y^n)\nonumber\\
	&\qquad\qquad +I(X^n;F_{\text{u}_2}|F_{\text{v}_1},F_{\text{u}_1},F_{\text{v}_2},Y^n)\nonumber\\
	&\leq 4\epsilon_n\label{eq:XnYnandFareind}
	\end{align}
	since 1) by (\ref{eq:independenceofFv1}) $F_{\text{v}_1}$ is almost independent of $(X^n,Y^n)$; 2) by (\ref{eq:independenceofFu1}) $F_{\text{u}_1}$ is almost independent of $(V_1^n,X^n,Y^n)$ and because $V_1^n$ determines $F_{\text{v}_1}$; 3) by (\ref{eq:independenceofFv2}) $F_{\text{v}_2}$ is almost independent of $(U_1^n,V_1^n,X^n,Y^n)$ and because $(V_1^n,U_1^n)$ determine $(F_{\text{v}_1},F_{\text{u}_1})$; 4) by (\ref{eq:independenceofFu2}) $F_{\text{u}_2}$ is almost independent of $(V_2^n,U_1^n,V_1^n,X^n,Y^n)$ and because $(V_1^n,U_1^n,V_2^n)$ determine $(F_{\text{v}_1},F_{\text{u}_1},F_{\text{v}_2})$;\\
	$(b)$ follows because $(V_1^n,V_2^n,U_1^n,U_2^n)$ determine $(W_1,W_2,F_1,F_2)$ and from the Markov chains $V_1^n-U_1^n-(X^n,Y^n,U_2^n,V_2^n)$ and $V_2^n-U_2^n-(X^n,Y^n,U_1^n)$;\\
	$(c)$ follows because $(X^n,U_1^n,U_2^n,Y^n)$ are i.i.d.
	
	\textbf{Privacy Leakage to Eve}: We have
	\begin{align}
	&I(X^n;W_1,W_2,F_1,F_2|Z^n)\nonumber\\
	&\overset{(a)}{=}H(W_1,W_2,F_1,F_2|Z^n)-H(W_1,W_2,F_1,F_2|X^n)\nonumber\\
	&\overset{(b)}{=} H(W_1,W_2,F_1,F_2|Z^n)-H(W_{u_1},F_{u_1},W_{u_2},F_{u_2},V_1^n,V_2^n|X^n)\nonumber\\
	&\qquad + H(V_1^n|W_1,W_2,F_1,F_2,X^n)+ H(V_2^n|V_1^n,W_1,W_2,F_1,F_2,X^n)\nonumber\\
	&\overset{(c)}{\leq} H(W_1,W_2,F_1,F_2|Z^n)-H(W_{u_1},F_{u_1},W_{u_2},F_{u_2},V_1^n,V_2^n|X^n)+2n\epsilon^{\prime}_n\nonumber\\
	&\overset{(d)}{=} H(W_1,W_2,F_1,F_2|Z^n) - H(U_1^n,U_2^n,V_1^n,V_2^n|X^n)\nonumber\\
	&\qquad+H(U_1^n|W_{u_1},F_{u_1},W_{u_2},F_{u_2},V_1^n,V_2^n,X^n)\nonumber\\
	&\qquad +H(U_2^n|U_1^n,W_{u_1},F_{u_1},W_{u_2},F_{u_2},V_1^n,V_2^n,X^n)+2n\epsilon^{\prime}_n\nonumber\\
	&\overset{(e)}{\leq} H(W_1,W_2,F_1,F_2|Z^n)\!-\!H(U_1^n,U_2^n,V_1^n,V_2^n|X^n)\!+\!4n\epsilon^{\prime}_n\nonumber\\
	&\overset{(f)}{=} H(W_1,W_2,F_1,F_2|Z^n) -nH(U_1,U_2,V_1,V_2|X)+4n\epsilon^{\prime}_n\label{eq:ach_privtoEvefirststep}
	\end{align}
	where $(a)$ follows because $(W_1,W_2,F_1,F_2)-X^n-Z^n$ form a Markov chain, $(b)$ follows since $(V_1^n,V_2^n)$ determine $(F_{v_1},W_{v_1},F_{v_2},W_{v_2})$, $(c)$ follows for some $\epsilon^{\prime}_n>0$ such that $\epsilon^{\prime}_n\rightarrow0$ when $n\rightarrow\infty$ because $(F_{v_1},W_{v_1},X^n)$ can reliably recover $V_1^n$ by (\ref{eq:Step1reconstr}), and similarly because $(F_{v_2},W_{v_2},V_1^n,X^n)$ can reliably recover $V_2^n$ by (\ref{eq:Step2reconstr}) both due to the Markov chain $(V_1^n,V_2^n)-X^n-Y^n$, $(d)$ follows because $(U_1^n,U_2^n)$ determine $(F_{u_1},W_{u_1},F_{u,2},W_{u_2})$, $(e)$ follows because $(F_{u_1},W_{u_1},V_1^n,V_2^n,X^n)$ can reliably recover $U_1^n$ by (\ref{eq:Step3reconstr}) and the inequality
	\begin{align}
	H(U_1|V_1,V_2,Y)\geq H(U_1|V_1,V_2,X)
	\end{align}  
	that follows from
	\begin{align}
	& I(U_1;V_1,V_2,X)-I(U_1;V_1,V_2,Y)\geq I(U_1;V_1,V_2,X)-I(U_1;V_1,V_2,Y,X)= 0\label{eq:proofthatHU1V1V2YisgreaterthanHU1V1V2X}
	\end{align}
	since $U_1-(V_1,V_2,X)-Y$ form a Markov chain. Furthermore, $(F_{u_2},W_{u_2},V_1^n,V_2^n,U_1^n,X^n)$ can reliably recover $U_2^n$ by (\ref{eq:Step4reconstr}) and the inequality 
	\begin{align}
	H(U_2|V_1,V_2,U_1,Y)\geq H(U_2|V_1,V_2,U_1,X)
	\end{align}
	that can be proved entirely similarly to (\ref{eq:proofthatHU1V1V2YisgreaterthanHU1V1V2X}) by using the Markov chain $U_2-(V_1,V_2,U_1,X)-Y$, and $(f)$ follows because $(U_1^n,U_2^n,V_1^n,V_2^n,X^n)$ are i.i.d. 
	
	In (\ref{eq:ach_privtoEvefirststep}), obtaining single letter bounds on the term $H(W_1,W_2,F_1,F_2|Z^n)$ requires analysis of numerous decodability cases, whereas there are only six different decodability cases analyzed in \cite{ourISITSecurePrivateFunctArxiv} for secure function computation with a single transmitting node. To simplify our analysis by applying the results in \cite{ourISITSecurePrivateFunctArxiv}, we combine the decoding order Steps~\ref{decorder:Step1} and \ref{decorder:Step2} given above such that $(V_1,V_2)$ are treated jointly and, similarly, we combine Steps~\ref{decorder:Step3} and \ref{decorder:Step4} such that $(U_1,U_2)$ are treated jointly. Using the combined steps, we can consider the six decodability cases analyzed in \cite[Section V-A]{ourISITSecurePrivateFunctArxiv} by replacing $V^n$ with $(V_1^n,V_2^n)$ and $U^n$ with $(U_1^n,U_2^n)$, respectively, in the proof. Since in (\ref{eq:ach_privtoEvefirststep}) the second term $-nH(U_1,U_2,V_1,V_2|X)$ can be obtained by applying the same replacement to the second term in \cite[Eq. (54)]{ourISITSecurePrivateFunctArxiv}, we obtain from (\ref{eq:ach_privtoEvefirststep}) and these decodability analyses that
	\begin{align}
	&I(X^n;W_1,W_2,F_1,F_2|Z^n)\nonumber\\
	&\leq n([I(U_1,U_2;Z|V_1,V_2)-I(U_1,U_2;Y|V_1,V_2)+\epsilon]^-\nonumber\\
	&\qquad\qquad +I(U_1,U_2;X|Z)+4\epsilon^{\prime}_n+\epsilon^{\prime\prime}_n)\label{eq:privEveachfinal}
	\end{align}
	for some $\epsilon^{\prime\prime}_n>0$ such that $\epsilon^{\prime\prime}_n\rightarrow0$ when $n\rightarrow\infty$.
	
	\textbf{Secrecy Leakage (to Eve)}: We obtain
	\begin{align}
	&I(\widetilde{X}^n_1,\widetilde{X}^n_2,Y^n;W_1,W_2,F_1,F_2|Z^n)\nonumber\\
	&\overset{(a)}{=}H(W_1,W_2,F_1,F_2|Z^n)-H(W_1,W_2,F_1,F_2|\widetilde{X}_1^n,\widetilde{X}_2^n)\nonumber\\
	&\overset{(b)}{=} H(W_1,W_2,F_1,F_2|Z^n) -H(W_{u_1},W_{u_2},F_{u_1},F_{u_2},V_1^n,V_2^n|\widetilde{X}_1^n,\widetilde{X}_2^n)\nonumber\\
	&\qquad +H(V_1^n|W_1,W_2,F_1,F_2,\widetilde{X}_1^n,\widetilde{X}_2^n) +H(V_2^n|V_1^n,W_1,W_2,F_1,F_2,\widetilde{X}_1^n,\widetilde{X}_2^n)\nonumber\\
	&\overset{(c)}{\leq} H(W_1,W_2,F_1,F_2|Z^n)-H(W_{u_1},W_{u_2},F_{u_1},F_{u_2},V_1^n,V_2^n|\widetilde{X}_1^n,\widetilde{X}_2^n)+2n\epsilon^{\prime}_n\nonumber\\
	&\overset{(d)}{=} H(W_1,W_2,F_1,F_2|Z^n) -H(U_1^n,U_2^n,V_1^n,V_2^n|\widetilde{X}_1^n,\widetilde{X}_2^n)+2n\epsilon^{\prime}_n\nonumber\\
	&\qquad + H(U_1^n|W_{u_1},W_{u_2},F_{u_1},F_{u_2},V_1^n,V_2^n,\widetilde{X}_1^n,\widetilde{X}_2^n)\nonumber\\
	&\qquad + H(U_2^n|U_1^n,W_{u_1},W_{u_2},F_{u_1},F_{u_2},V_1^n,V_2^n,\widetilde{X}_1^n,\widetilde{X}_2^n)\nonumber\\
	&\overset{(e)}{\leq} H(W_1,W_2,F_1,F_2|Z^n)-H(U_1^n,U_2^n,V_1^n,V_2^n|\widetilde{X}_1^n,\widetilde{X}_2^n)+4n\epsilon^{\prime}_n\nonumber\\
	&\overset{(f)}{\leq}H(W_1,W_2,F_1,F_2|Z^n) -nH(U_1,U_2,V_1,V_2|\widetilde{X}_1,\widetilde{X}_2)+4n\epsilon^{\prime}_n\label{eq:ach_secrecyfirststep}
	\end{align}
	where $(a)$ follows from the Markov chain $(W_1,W_2,F_1,F_2)-(\widetilde{X}_1^n,\widetilde{X}_2^n)-(Y^n,Z^n)$, $(b)$ follows since $(V_1^n,V_2^n)$ determine $(F_{v_1},W_{v_1},F_{v_2},W_{v_2})$, $(c)$ follows because $(F_{v_1},W_{v_1},\widetilde{X}_1^n,\widetilde{X}_2^n)$ can reliably recover $V_1^n$ by (\ref{eq:Step1reconstr}), and similarly because $(F_{v_2},W_{v_2},V_1^n,\widetilde{X}_1^n,\widetilde{X}_2^n)$ can reliably recover $V_2^n$ by (\ref{eq:Step2reconstr}) both due to the Markov chain $(V_1^n,V_2^n)-(\widetilde{X}_1^n,\widetilde{X}_2^n)-Y^n$, $(d)$ follows since $(U_1^n,U_2^n)$ determine $(F_{u_1},W_{u_1},F_{u_2},W_{u_2})$, $(e)$ follows because $(F_{u_1},W_{u_1},V_1^n,V_2^n,\widetilde{X}_1^n,\widetilde{X}_2^n)$ can reliably recover $U_1^n$ by (\ref{eq:Step3reconstr}) and the inequality 
	\begin{align}
	H(U_1|V_1,V_2,Y)\geq H(U_1|V_1,V_2,\widetilde{X}_1,\widetilde{X}_2^n)
	\end{align}
	that can be proved similarly to (\ref{eq:proofthatHU1V1V2YisgreaterthanHU1V1V2X}) due to the Markov chain $U_1-(V_1,V_2,\widetilde{X}_1,\widetilde{X}_2)-Y$. Furthermore, $(F_{u_2},W_{u_2},V_1^n,V_2^n,U_1^n,\widetilde{X}_1^n,\widetilde{X}_2^n)$ can reliably recover $U_2^n$ by (\ref{eq:Step4reconstr}) and the inequality 
	\begin{align}
	H(U_2|V_1,V_2,U_1,Y)\geq H(U_2|V_1,V_2,U_1,\widetilde{X}_1,\widetilde{X}_2)
	\end{align}
	that can be proved by using the Markov chain $U_2-(V_1,V_2,U_1,\widetilde{X}_1,\widetilde{X}_2)-Y$, and $(f)$ follows because $(U_1^n,U_2^n,V_1^n,V_2^n,\widetilde{X}_1^n,\widetilde{X}_2^n)$ are i.i.d. 
	
	We remark that the terms in (\ref{eq:ach_secrecyfirststep}) are entirely similar to the terms in (\ref{eq:ach_privtoEvefirststep}). One can show that all steps of the decodability analysis from \cite[Section~V-A]{ourISITSecurePrivateFunctArxiv} that is applied to (\ref{eq:ach_privtoEvefirststep}) can be applied also to (\ref{eq:ach_secrecyfirststep}) by replacing $X$ with $(\widetilde{X}_1,\widetilde{X}_2)$, so we obtain
	\begin{align}
	&I(\widetilde{X}^n_1,\widetilde{X}^n_2,Y^n;W_1,W_2,F_1,F_2|Z^n)\nonumber\\
	&\leq n[I(U_1,U_2;Z|V_1,V_2)-I(U_1,U_2;Y|V_1,V_2)+\epsilon]^-\nonumber\\
	&\qquad +nI(U_1,U_2;\widetilde{X}_1,\widetilde{X}_2|Z)+5n\epsilon^{\prime}_n \label{eq:ach_secrecyleakCase5}.
	\end{align}
	
	We consider that the public indices $(F_1,F_2)$ are generated uniformly at random and the encoders generate $(V_1^n,U_1^n)$ and $(V_2^n,U_2^n)$ according to $P_{V_1^nU_1^nV_2^nU_2^n|\widetilde{X}_1^nF_{1}\widetilde{X}_2^nF_{2}}$ obtained from the binning scheme above. This procedure induces a joint probability distribution that is almost equal to $P_{V_1V_2U_1U_2\widetilde{X}_1\widetilde{X}_2XYZ}$ fixed as in (\ref{eq:iid}) \cite[Section 1.6]{BlochLectureNotes2018}. Since the privacy and secrecy leakage metrics considered above are expectations over all possible realizations $F=f$, applying the selection lemma \cite[Lemma~2.2]{Blochbook}, these results prove the achievability for Theorem~\ref{theo:innerouterfor2nodeslossless} by choosing an $\epsilon>0$ such that $\epsilon\rightarrow 0$ when $n\rightarrow\infty$. We remark that the achievable region is convexified by using a time-sharing random variable $Q$ such that $P_{QV_1V_2}=P_QP_{V_1|Q}P_{V_2|Q}$, required because of the $[\cdot]^-$ operation. 
\end{proof}

\subsection{Outer Bound}\label{subsec:outerboundproof2lossless}
\begin{proof}[Proof Sketch]
	Assume that for some $n\geq 1$ and $\delta_n\!>\!0$, there exist two encoders and a decoder such that (\ref{eq:reliability_cons})-(\ref{eq:privEve_cons}) are satisfied for some tuple $(R_\text{s}, R_{\text{w}_1},R_{\text{w},2},R_{\ell,{\text{Dec}}}, R_{\ell,{\text{Eve}}})$. Let 
	\begin{align}
	&V_{1,i}\triangleq (W_1,Y^{n}_{i+1},Z^{i-1})\label{eq:defV1}\\ 
	&V_{2,i}\triangleq (W_2,Y^{n}_{i+1},Z^{i-1})\label{eq:defV2}\\
	&U_{1,i}\triangleq (X^{i-1},W_1,Y^{n}_{i+1},Z^{i-1})\label{eq:defU1}\\ 
	&U_{2,i}\triangleq (X^{i-1},W_2,Y^{n}_{i+1},Z^{i-1})\label{eq:defU2}
	\end{align}
	that satisfy the Markov chains
	\begin{align}
	&V_{1,i}-U_{1,i}-\widetilde{X}_{1,i}-X_i-(\widetilde{X}_{2,i},Y_{i},Z_{i})\label{eq:MarkovV1U1others}\\
	&V_{2,i}-U_{2,i}-\widetilde{X}_{2,i}-X_i-(\widetilde{X}_{1,i},Y_{i},Z_{i})\label{eq:MarkovV2U2others}.
	\end{align}

	\textbf{Admissibility of $(\mathbf{U_1},\mathbf{U_2})$}: Define 
	\begin{align}
	n\epsilon_n=n\delta_n \log(|\mathcal{\widetilde{X}}_1||\mathcal{\widetilde{X}}_2||\mathcal{Y}|) + H_b(\delta_n)
	\end{align}
	such that $\epsilon_n\!\rightarrow\!0$ if $\delta_n\!\rightarrow\!0$. Using Fano's inequality and (\ref{eq:reliability_cons}), we obtain
	\begin{align}
	&n\epsilon_n\geq H(f^n|\widehat{f^n})\nonumber\\
	&\overset{(a)}{=}H(f^n|\widebar{f}^n)=\sum_{i=1}^nH(f_i|\widebar{f}_i)\nonumber\\
	&\geq\sum_{i=1}^nH(f_i|\widebar{f}^n)\overset{(b)}{\geq}\sum_{i=1}^n H(f_i|W_1,W_2,Y^n)\nonumber\\
	&\geq \sum_{i=1}^nH(f_i|W_1,W_2,Y^n,X^{i-1}, Z^{i-1})\nonumber\\
	&\overset{(c)}{=}\sum_{i=1}^nH(f_i|W_1,W_2,Y^n_{i+1},X^{i-1}, Z^{i-1},Y_i)\nonumber\\
	&\overset{(d)}{=}\sum_{i=1}^nH(f_i|U_{1,i},U_{2,i},Y_i)\label{eq:admissibilityandFano} 
	\end{align}
	where $(a)$ follows from \cite[Lemma 2]{CorrelatedPaperLong} that proves that when $n\rightarrow\infty$, there exists an i.i.d. random variable $\widebar{f}^n$ that satisfies both
	\begin{align}
	H(f^n|\widehat{f^n})=H(f^n|\widebar{f}^n)
	\end{align}
	and the Markov chain 
	\begin{align}
	\widehat{f^n}-\widebar{f}^n-(W_1,W_2,Y^n)
	\end{align}
	$(b)$ follows from the data processing inequality because of the Markov chain 
	\begin{align}
	f^n-(W_1,W_2,Y^n)-\widebar{f}^n
	\end{align}
	and permits randomized decoding, $(c)$ follows from the Markov chain
	\begin{align}
	Y^{i-1}-(X^{i-1},Z^{i-1},W_1,W_2,Y_i,Y_{i+1}^n)-f_i \label{eq:conv_Markov2}
	\end{align}
	and $(d)$ follows from the definitions of $U_{1,i}$ and $U_{2,i}$.

	\textbf{Public Message (Storage) Rates}: We obtain
	\begin{align}
	&n(R_{\text{w}_1}+\delta_n) \overset{(a)}{\geq} \log|\mathcal{W}_1| \nonumber\\
	&\geq H(W_1|Y^n)-H(W_1|\widetilde{X}_1^n,Y^n)\nonumber\\
	&= H(\widetilde{X}_1^n|Y^n)-H(\widetilde{X}_1^n|W_1,Y^n)\nonumber\\
	& =H(\widetilde{X}_1^n|Y^n)-\sum_{i=1}^nH(\widetilde{X}_{1,i}|\widetilde{X}_1^{i-1},W_1,Y^n)\nonumber\\
	&\overset{(b)}{=}H(\widetilde{X}_1^n|Y^n)-\sum_{i=1}^nH(\widetilde{X}_{1,i}|\widetilde{X}_1^{i-1},W_1,Y_{i+1}^n,Y_i)\nonumber
					\end{align}
	\begin{align}
	&\overset{(c)}{\geq}H(\widetilde{X}_1^n|Y^n)-\sum_{i=1}^nH(\widetilde{X}_{1,i}|X^{i-1},Z^{i-1},W_1,Y_{i+1}^n,Y_i)\nonumber\\
	&\overset{(d)}{=}nH(\widetilde{X}_{1}|Y)-\sum_{i=1}^nH(\widetilde{X}_{1,i}|U_{1,i},Y_i) \nonumber\\
	&=\sum_{i=1}^n I(U_{1,i};\widetilde{X}_{1,i}|Y_i)\nonumber\\
	&\overset{(e)}{=} \sum_{i=1}^n [I(V_{1,i};\widetilde{X}_{1,i}|Y_i)+I(U_{1,i};\widetilde{X}_{1,i}|Y_i,V_{1,i})]\nonumber\\
	&= \sum_{i=1}^n \Big[I(V_{1,i};\widetilde{X}_{1,i},V_{2,i}|Y_i)-I(V_{1,i};V_{2,i}|\widetilde{X}_{1,i},Y_i)+I(U_{1,i};\widetilde{X}_{1,i},U_{2,i}|Y_i,V_{1,i})\nonumber\\
	&\qquad\qquad- I(U_{1,i};U_{2,i}|\widetilde{X}_{1,i},Y_i,V_{1,i})\Big]\nonumber\\
	&\geq \sum_{i=1}^n \Big[I(V_{1,i};\widetilde{X}_{1,i}|Y_i,V_{2,i})-I(V_{1,i};V_{2,i}|\widetilde{X}_{1,i},Y_i) +I(U_{1,i};\widetilde{X}_{1,i}|Y_i,V_{1,i},U_{2,i})\nonumber\\
	&\qquad\qquad- I(U_{1,i};U_{2,i}|\widetilde{X}_{1,i},Y_i,V_{1,i})\Big]\label{eq:storagerate1conv}
	\end{align}
	where $(a)$ follows by (\ref{eq:storage_cons1}), $(b)$ follows from the Markov chain
	\begin{align}
	Y^{i-1}-(\widetilde{X}_1^{i-1},W_1,Y_{i+1}^n,Y_i)-\widetilde{X}_{1,i}\label{eq:markovyiminus1xtildei}
	\end{align}
	$(c)$ follows from the data processing inequality applied to the Markov chain
	\begin{align}
	&(X^{i-1},Z^{i-1})-(\widetilde{X}_1^{i-1}, W_1,Y_{i+1}^n,Y_i)-\widetilde{X}_{1,i}\label{eq:Markovconversextildeandximinus1}
	\end{align}
	$(d)$ follows from the definition of $U_{1,i}$, and $(e)$ follows by (\ref{eq:MarkovV1U1others}). Similarly, one can show by symmetry that we have
	\begin{align}
	&n(R_{\text{w}_2}+\delta_n)\nonumber\\
	&\geq\sum_{i=1}^n \Big[I(V_{2,i};\widetilde{X}_{2,i}|Y_i,V_{1,i})-I(V_{2,i};V_{1,i}|\widetilde{X}_{2,i},Y_i)\nonumber\\
	&\qquad\qquad +I(U_{2,i};\widetilde{X}_{2,i}|Y_i,V_{2,i},U_{1,i})- I(U_{2,i};U_{1,i}|\widetilde{X}_{2,i},Y_i,V_{2,i})\Big]\label{eq:storagerate2conv}.
	\end{align}
	
	Now we consider the sum-rate bound such that	
	\begin{align}
&n(R_{\text{w}_1}+\delta_n)+n(R_{\text{w}_2}+\delta_n) \overset{(a)}{\geq} \log(|\mathcal{W}_1|\cdot|\mathcal{W}_2|)\nonumber\\
&\geq H(W_1,W_2|Y^n)-H(W_1,W_2|\widetilde{X}_1^n,\widetilde{X}_2^n,Y^n)\nonumber\\
&\overset{(b)}{=}\sum_{i=1}^n \Big[H(\widetilde{X}_{1,i},\widetilde{X}_{2,i}|Y_i) -H(\widetilde{X}_{1,i},\widetilde{X}_{2,i} |\widetilde{X}^{i-1}_1,\widetilde{X}^{i-1}_2,Y_{i}^n,W_1,W_2)\Big]\nonumber\\
&\overset{(c)}{\geq}\sum_{i=1}^n \Big[H(\widetilde{X}_{1,i},\widetilde{X}_{2,i}|Y_i) -H(\widetilde{X}_{1,i},\widetilde{X}_{2,i} |X^{i-1},Z^{i-1},Y_{i}^n,W_1,W_2)\Big]\nonumber
						\end{align}
	\begin{align}
	&\overset{(d)}{=}\sum_{i=1}^n I(U_{1,i},U_{2,i};\widetilde{X}_{1,i},\widetilde{X}_{2,i}|Y_i)\nonumber\\
	&\overset{(e)}{=}\sum_{i=1}^n \Big[I(U_{1,i},U_{2,i};\widetilde{X}_{1,i},\widetilde{X}_{2,i}|Y_i,V_{1,i},V_{2,i}) +I(V_{1,i},V_{2,i};\widetilde{X}_{1,i},\widetilde{X}_{2,i}|Y_i)\Big]\nonumber\\
	& \overset{(f)}{=}\sum_{i=1}^n \Big[I(U_{1,i};\widetilde{X}_{1,i},\widetilde{X}_{2,i}|Y_i,V_{1,i},V_{2,i})+ I(U_{2,i};\widetilde{X}_{1,i},\widetilde{X}_{2,i}|Y_i,U_{1,i},V_{2,i})\nonumber\\
	&\qquad\qquad +I(V_{1,i};\widetilde{X}_{1,i},\widetilde{X}_{2,i}|Y_i) +I(V_{2,i};\widetilde{X}_{1,i},\widetilde{X}_{2,i}|Y_i,V_{1,i})\Big]\nonumber\\
	&\geq \sum_{i=1}^n \Big[I(U_{1,i};\widetilde{X}_{1,i}|Y_i,V_{1,i},V_{2,i})+ I(U_{2,i};\widetilde{X}_{2,i}|Y_i,U_{1,i},V_{2,i})\nonumber\\
	&\qquad\qquad +I(V_{1,i};\widetilde{X}_{1,i}|Y_i)+I(V_{2,i};\widetilde{X}_{2,i}|Y_i,V_{1,i})\Big]
	\end{align}	
	where $(a)$ follows by (\ref{eq:storage_cons1}) and (\ref{eq:storage_cons2}), $(b)$ follows since $(\widetilde{X}_1^n,\widetilde{X}_2^n,Y^n)$ are i.i.d. and because
	\begin{align}
	Y^{i-1}-(\widetilde{X}_1^{i-1},\widetilde{X}_2^{i-1},W_1,W_2,Y_i^n)-(\widetilde{X}_{1,i},\widetilde{X}_{2,i})\label{eq:MarkovYiminus1Xtilde}
	\end{align}
	form a Markov chain, $(c)$ follows by applying the data processing inequality to the Markov chain
	\begin{align}
	(X^{i-1},Z^{i-1})-(\widetilde{X}_1^{i-1},\widetilde{X}_2^{i-1},W_1,W_2,Y_i^n)-(\widetilde{X}_{1,i},\widetilde{X}_{2,i})\label{eq:MarkovXZiminus1Xtilde}
	\end{align}
	$(d)$ follows from the definitions of $U_{1,i}$ and $U_{2,i}$, $(e)$ follows from the Markov chain
	\begin{align}
	(V_{1,i},V_{2,i})-(U_{1,i},U_{2,i})-(\widetilde{X}_{1,i},\widetilde{X}_{2,i})-Y_i\label{eq:MarkovV12U12Xtilde12Y}
	\end{align}
	and $(f)$ follows from the Markov chain
	\begin{align}
	V_{1,i}-(U_{1,i},Y_i,V_{2,i})-(U_{2,i},\widetilde{X}_{1,i},\widetilde{X}_{2,i}).
	\end{align}
	
	\textbf{Privacy Leakage to Decoder}: We have
	\begin{align}
	&n(R_{\ell,\text{Dec}}+\delta_n)\\ &\overset{(a)}{\geq}H(W_1,W_2|Y^n)-H(W_1,W_2|X^n)\nonumber\\
	&\overset{(b)}{=}\sum_{i=1}^n \Big[I(W_1,W_2;X_i|X^{i-1},Y_{i+1}^n)  - I(W_1,W_2;Y_i|Y_{i+1}^n,X^{i-1})\Big]\nonumber\\
	&\overset{(c)}{=}\sum_{i=1}^n \Big[I(W_1,W_2;X_i|X^{i-1},Z^{i-1},Y_{i+1}^n) - I(W_1,W_2;Y_i|Y_{i+1}^n,X^{i-1},Z^{i-1})\Big]\nonumber\\
	&\overset{(d)}{=}\sum_{i=1}^n \Big[I(W_1,W_2,X^{i-1},Z^{i-1},Y_{i+1}^n;X_i)- I(W_1,W_2,Y_{i+1}^n,X^{i-1},Z^{i-1};Y_i)\Big]\nonumber\\
	&\overset{(e)}{=}\sum_{i=1}^n \Big[I(U_{1,i},U_{2,i};X_i) - I(U_{1,i},U_{2,i};Y_i)\Big]\nonumber\\
	&\overset{(f)}{=}\sum_{i=1}^n  I(U_{1,i},U_{2,i};X_i|Y_i)\label{eq:privacytoDecconverselossless} 
	\end{align}
	where $(a)$ follows by (\ref{eq:privDec_cons}) and from the Markov chain $(W_1,W_2)-X^n-Y^n$, $(b)$ follows from Csisz\'{a}r's sum identity, $(c)$ follows from the Markov chain
	\begin{align}
	&Z^{i-1}- (X^{i-1},Y_{i+1}^n)-(X_i,Y_i,W_1,W_2)\label{eq:MarkovZiminusXiminus1}
	\end{align}
	$(d)$ follows because $(X^n,Y^n,Z^n)$ are i.i.d., $(e)$ follows from the definitions of $U_{1,i}$ and $U_{2,i}$, and $(f)$ follows from the Markov chain 
	\begin{align}
	(U_{1,i},U_{2,i})-X_i-Y_i.
	\end{align}

	\textbf{Privacy Leakage to Eve}: We have
	\begin{align}
	&n(R_{\ell,\text{Eve}}+\delta_n)\nonumber\\ 
	&\overset{(a)}{\geq}[H(W_1,W_2|Z^n)-H(W_1,W_2|Y^n)]+[H(W_1,W_2|Y^n)-H(W_1,W_2|X^n)]\nonumber\\
	&\overset{(b)}{=}\sum_{i=1}^n\Big[I(W_1,W_2;Y_i|Y_{i+1}^n,Z^{i-1}) -I(W_1,W_2;Z_i|Z^{i-1},Y_{i+1}^n)\Big]\nonumber\\
	&\qquad+ \sum_{i=1}^n\Big[I(W_1,W_2;X_i|X^{i-1},Y_{i+1}^n) -I(W_1,W_2;Y_i|Y_{i+1}^n,X^{i-1})\Big]\nonumber\\
	&\overset{(c)}{=}\sum_{i=1}^n\Big[I(W_1,W_2;Y_i|Y_{i+1}^n,Z^{i-1}) -I(W_1,W_2;Z_i|Z^{i-1},Y_{i+1}^n)\Big]\nonumber\\
	&\qquad+ \sum_{i=1}^n\Bigg[I(W_1,W_2;X_i|X^{i-1},Y_{i+1}^n,Z^{i-1}) -I(W_1,W_2;Y_i|Y_{i+1}^n,X^{i-1},Z^{i-1})\Bigg]\nonumber\\
	&\overset{(d)}{=}\sum_{i=1}^n\Big[I(W_1,W_2,Y_{i+1}^n,Z^{i-1};Y_i)-I(W_1,W_2,Z^{i-1},Y_{i+1}^n;Z_i)\Big]\nonumber\\
	&\qquad+\sum_{i=1}^n\Bigg[I(W_1,W_2,X^{i-1},Y_{i+1}^n,Z^{i-1};X_i) -I(W_1,W_2,Y_{i+1}^n,X^{i-1},Z^{i-1};Y_i)\Bigg]\nonumber\\
	&\overset{(e)}{=} \sum_{i=1}^n \Big[I(V_{1,i},V_{2,i};Y_i)-I(V_{1,i},V_{2,i};Z_i) +I(U_{1,i},U_{2,i}V_{1,i},V_{2,i};X_i)\nonumber\\
	&\qquad\qquad-I(U_{1,i},U_{2,i},V_{1,i},V_{2,i};Y_i)\Big]\nonumber\\
	&= \sum_{i=1}^n \Big[-I(U_{1,i},U_{2,i},V_{1,i},V_{2,i};Z_i) +I(U_{1,i},U_{2,i},V_{1,i},V_{2,i};X_i)\nonumber\\
	&\qquad\qquad+I(U_{1,i},U_{2,i};Z_i|V_{1,i},V_{2,i})-I(U_{1,i},U_{2,i};Y_i|V_{1,i},V_{2,i})\Big]\nonumber\\
	&\overset{(f)}{\geq}\! \sum_{i=1}^n\Bigg[I(U_{1,i},U_{2,i};X_i|Z_i)+\Big[I(U_{1,i},U_{2,i};Z_i|V_{1,i},V_{2,i})-I(U_{1,i},U_{2,i};Y_i|V_{1,i},V_{2,i})\Big]^-\Bigg]
	\end{align}
	where $(a)$ follows by (\ref{eq:privEve_cons}) and from the Markov chain $(W_1,W_2)-X^n-Z^n$, $(b)$ follows from Csisz\'{a}r's sum identity, $(c)$ follows from the Markov chain in (\ref{eq:MarkovZiminusXiminus1}), $(d)$ follows because $(X^n,Y^n,Z^n)$ are i.i.d., $(e)$ follows from the definitions of $V_{1,i}$, $V_{2,i}$, $U_{1,i}$ and $U_{2,i}$, and $(f)$ follows from the Markov chain 
	\begin{align}
	(V_{1,i},V_{2,i})-(U_{1,i},U_{2,i})-X_i-Z_i.
	\end{align}

	\textbf{Secrecy Leakage (to Eve)}: We obtain
	\begin{align}
	&n(R_{\text{s}}+\delta_n)\nonumber\\ &\overset{(a)}{\geq}[H(W_1,W_2|Z^n)-H(W_1,W_2|Y^n)] +[H(W_1,W_2|Y^n)-H(W_1,W_2|\widetilde{X}_1^n,\widetilde{X}_2^n,Y^n)]\nonumber\\
	&\overset{(b)}{=}\sum_{i=1}^n\Big[I(W_1,W_2;Y_i|Y_{i+1}^n,Z^{i-1}) -I(W_1,W_2;Z_i|Z^{i-1},Y_{i+1}^n)\nonumber\\
	&\qquad\qquad+H(\widetilde{X}_{1,i},\widetilde{X}_{2,i}|Y_i)-\!H(\widetilde{X}_{1,i},\widetilde{X}_{2,i}|\widetilde{X}_1^{i-1},\widetilde{X}_2^{i-1},W_1,W_2,Y_{i+1}^n,Y_i)\Big]\nonumber\\
	&\overset{(c)}{\geq}\sum_{i=1}^n\Big[I(W_1,W_2,Y_{i+1}^n,Z^{i-1};Y_i) -I(W_1,W_2,Z^{i-1},Y_{i+1}^n;Z_i)\nonumber\\
	&\qquad\qquad+H(\widetilde{X}_{1,i},\widetilde{X}_{2,i}|Y_i)-H(\widetilde{X}_{1,i},\widetilde{X}_{2,i}|X^{i-1},Z^{i-1},W_1,W_2,Y_{i+1}^n,Y_i)\Big]\nonumber\\
	&\overset{(d)}{=} \sum_{i=1}^n\Big[I(V_{1,i},V_{2,i};Y_i)-I(V_{1,i},V_{2,i};Z_i) +I(U_{1,i},U_{2,i},V_{1,i},V_{2,i};\widetilde{X}_{1,i},\widetilde{X}_{2,i}|Y_i)\Big]\nonumber\\
	&\overset{(e)}{=} \sum_{i=1}^n
	\Big[I(V_{1,i},V_{2,i};Y_i)-I(V_{1,i},V_{2,i};Z_i)\nonumber\\
	&\qquad\qquad+I(U_{1,i},U_{2,i},V_{1,i},V_{2,i};\widetilde{X}_{1,i},\widetilde{X}_{2,i})-I(U_{1,i},U_{2,i},V_{1,i},V_{2,i};Y_i)\Big]\nonumber\\
	&= \sum_{i=1}^n
	\Big[-\!I(U_{1,i},U_{2,i},V_{1,i},V_{2,i};Z_i) +I(U_{1,i}U_{2,i},V_{1,i},V_{2,i};\widetilde{X}_{1,i},\widetilde{X}_{2,i})\nonumber\\
	&\qquad\qquad+I(U_{1,i},U_{2,i};Z_i|V_{1,i},V_{2,i})-I(U_{1,i},U_{2,i};Y_i|V_{1,i},V_{2,i})\Big]\nonumber\\
	&\overset{(f)}{\geq} \sum_{i=1}^n\Bigg[ I(U_{1,i},U_{2,i};\widetilde{X}_{1,i},\widetilde{X}_{2,i}|Z_i)+\Big[I(U_{1,i},U_{2,i};Z_i|V_{1,i},V_{2,i})-I(U_{1,i},U_{2,i};Y_i|V_{1,i},V_{2,i})\Big]^-\Bigg]
	\end{align}
	where $(a)$ follows by (\ref{eq:secrecyleakage_cons}), $(b)$ follows because $(\widetilde{X}_1^n,\widetilde{X}_2^n, Y^n)$ are i.i.d., and from Csisz\'{a}r's sum identity and the Markov chain in (\ref{eq:MarkovYiminus1Xtilde}), $(c)$ follows because $(Y^n,Z^n)$ are i.i.d. and from the data processing inequality applied to the Markov chain in (\ref{eq:MarkovXZiminus1Xtilde}), $(d)$ follows from the definitions of $V_{1,i}$, $V_{2,i}$, $U_{1,i}$, and $U_{2,i}$, $(e)$ follows from the Markov chain given in (\ref{eq:MarkovV12U12Xtilde12Y}), and $(f)$ follows from the Markov chain
	\begin{align}
	(V_{1,i},V_{2,i})-(U_{1,i},U_{2,i})-(\widetilde{X}_{1,i},\widetilde{X}_{2,i})-Z_i.
	\end{align}		
	
	Introduce a uniformly distributed time-sharing random variable $\displaystyle Q\!\sim\! \text{Unif}[1\!:\!n]$ that is independent of other random variables, and define $X\!=\!X_Q$, $\displaystyle \widetilde{X}_1\!=\!\widetilde{X}_{1,Q}$, $\displaystyle \widetilde{X}_2\!=\!\widetilde{X}_{2,Q}$, $\displaystyle Y\!=\!Y_Q$, $\displaystyle Z\!=\!Z_Q$, $V_1\!=\!V_{1,Q}$, $V_2\!=\!V_{2,Q}$, $U_1\!=\!(U_{1,Q},\!Q)$, $U_2\!=\!(U_{2,Q},\!Q)$, and $f=f_Q$, so
	\begin{align}
	& (Q,V_1)\!-U_1-\widetilde{X}_1-X-(\widetilde{X}_2,Y,Z)\\ 
	& (Q,V_2)\!-U_2-\widetilde{X}_2-X-(\widetilde{X}_1,Y,Z)
	\end{align}
	form Markov chains. The proof of the outer bound follows by letting $\delta_n\rightarrow0$.
	
	\textbf{Cardinality Bounds}: We use the support lemma \cite[Lemma 15.4]{CsiszarKornerbook2011} to prove the cardinality bounds and apply similar steps as in \cite{ourISITSecurePrivateFunctArxiv, LifengFCTrans}, so we omit the proof.
\end{proof}

\section{Conclusion}\label{sec:conclusion}
We considered the function computation problem, where three nodes observe correlated random variables and aim to compute a target function of their observations at the fusion center node. We modeled the source of the correlation between these nodes by positing that all three random variables are noisy observations of a remote random source. Furthermore, we imposed one secrecy, two privacy, and two storage constraints with operational meanings on this function computation problem to define a lossless rate region by considering an eavesdropper that observes a correlated random variable. The lossless function computation problem was extended
by allowing the function computed to be a distorted version of the target function, which defined the lossy function computation problem.

We proposed inner and outer bounds for the lossless and lossy rate regions. The secrecy leakage and privacy leakage rates that are measured with respect to the eavesdropper were shown to be different due to the remote source considered, unlike in the literature. Furthermore, we established simplified rate region bounds for functions that are partially invertible with respect to one of the transmitting node observations as well as for invertible functions. Moreover, we considered two different physical-degradation cases for the measurement channels of the eavesdropper and fusion center when the function computed was invertible. We derived the corresponding rate region bounds, one of which is evaluated as an example scenario.

In future work, we will propose inner and outer bounds for the lossless and lossy multi-function computation problems with multiple transmitting nodes.

\vspace{6pt} 




\funding{This research was supported by the German Federal Ministry of Education and Research (BMBF) within the national initiative for ``Post Shannon Communication (NewCom)'' under the Grant 16KIS1004 and by the German Research Foundation (DFG) under the Grant SCHA 1944/9-1.}

\acknowledgments{O. G{\"u}nl{\"u} thanks Matthieu Bloch and Rafael F. Schaefer for their contributions to the conference papers used in this work.}

\conflictsofinterest{The author declares no conflict of interest. The funders had no role in the design of the study; in the collection, analyses, or interpretation of data; in the writing of the manuscript, or in the decision to publish the results.}


\appendixtitles{no} 

\end{paracol}
\reftitle{References}


\externalbibliography{yes}
\bibliography{referencesEntropy.bib}


%


\end{document}